\documentclass[aps,superscriptaddress,amsmath,amssymb,twocolumn,amsfonts,floatfix, longbibliography]{revtex4-1}
\usepackage{mathtools}
\usepackage{braket}
\usepackage[dvipsnames]{xcolor}
\usepackage{float}
\usepackage{subfigure}
\usepackage{upgreek}
\usepackage{pgffor}
\usepackage{float}
\usepackage{silence}
\usepackage[colorlinks=true,linktoc=page,linkcolor=blue,citecolor=magenta,urlcolor=Fuchsia]{hyperref}
\usepackage[normalem]{ulem}
\usepackage{sidecap,tikz}
\usepackage{orcidlink}

\newcommand{\beq}{\begin{equation}}
\newcommand{\eeq}{\end{equation}}
\newcommand{\bea}{\begin{eqnarray}}
\newcommand{\eea}{\end{eqnarray}}

\newcommand{\eqn}[1] {Eq.~(\ref{#1})}
\newcommand{\fig}[1]{Fig.~\ref{#1}}
\newcommand{\sect}[1]{Sec.~\ref{#1}}

\mathchardef\mhyphen="2D % Define a "math hyphen"
\newcommand{\noin}{\noindent}

\newcommand{\ua}{{\uparrow }}
\newcommand{\da}{{\downarrow }}
\newcommand{\eps}{{\epsilon}}

\newcommand{\non}{\nonumber}

\begin{document}
%=============START of MAIN PAPER===============
% \title{Topological superconductivity and field-free superconducting diode effect in altermagnetic Shiba chains}	
% \title{Coexistence of Field-Free Superconducting Diode Effect and Topological Superconductivity in Altermagnetic Shiba Chains}
\title{Field-free Superconducting Diode Effect and Topological Fulde-Ferrell Superconductivity in Altermagnetic Shiba Chains}

\author{{Dibyendu Samanta}\,\orcidlink{0009-0004-3022-7633}}
\email{dibyendus21@iitk.ac.in}
\affiliation{Department of Physics, Indian Institute of Technology, Kanpur 208016, India}

\author{{Sudeep Kumar Ghosh}\,\orcidlink{0000-0002-3646-0629}}
\email{skghosh@iitk.ac.in}
\affiliation{Department of Physics, Indian Institute of Technology, Kanpur 208016, India}
	
\date{\today}
%-------------------------------------------------------

\begin{abstract}
The superconducting diode effect (SDE), characterized by a directional asymmetry in the critical supercurrents, typically requires external magnetic fields to break time-reversal symmetry-- posing challenges for scalability and device integration. Here, we demonstrate a field-free realization of the SDE in a helical Shiba chain proximitized by a $d$-wave altermagnet. Using a self-consistent Bogoliubov–de Gennes approach, we uncover a topological Fulde–Ferrell (FF) superconducting state that hosts tunable Majorana zero modes at the chain ends. The Cooper pair momentum is directly controlled by an externally injected supercurrent providing an experimentally accessible tuning parameter for driving and manipulating the topological FF phase. This state is stabilized by the interplay between the exchange coupling of magnetic adatoms and the induced altermagnetic spin splitting. Crucially, the same topological FF phase supports strong nonreciprocal supercurrents, achieving diode efficiencies exceeding $45\%$ without applied magnetic fields. The $d$-wave altermagnet plays a dual role: it intrinsically breaks time-reversal symmetry, enabling topological superconductivity, and introduces inversion symmetry breaking via momentum-dependent spin-splitting, driving the field-free SDE in a junction-free architecture. Our results establish the Shiba chain–altermagnet heterostructure as a promising platform for realizing topological superconducting devices with efficient, intrinsic superconducting diode functionality-- offering a scalable pathway towards dissipationless quantum technologies.

\end{abstract}
	
\maketitle

%--------------------------------------------------------
\section{Introduction}
%--------------------------------------------------------

Topological superconductors, capable of hosting Majorana zero modes (MZMs)~\cite{Kitaev_2001,Lutchyn_2010,Oreg_2010,Qi_2011,Alicea_2012,Leijnse_2012,Beenakker_2013}, have attracted significant interest due to their potential for fault-tolerant quantum computation via non-Abelian braiding statistics~\cite{Ivanov_2001,Kitaev_2003,stern2010non,Chetan_2008}. Following Kitaev's seminal p-wave model~\cite{Kitaev_2001}, experimental realizations have included Rashba nanowires  in proximity to s-wave superconductors under a Zeeman field~\cite{Lutchyn_2010,Oreg_2010} and quantum spin Hall edges coupled to superconductors~\cite{LiangFu_2008,LiangFu_2009}. More recently, Shiba chains~\cite{Rachel2025,Choi2019,Pientka_2013} constructed by placing magnetic adatoms on s-wave superconductors have emerged as promising platforms, where helical magnetic order and superconducting pairing create topological phases with end-localized MZMs through hybridized Yu-Shiba-Rusinov (YSR) states. However, such systems typically require external magnetic fields that can suppress superconductivity. This limitation has motivated exploration of heterostructures of superconductors with altermagnets~\cite{Ghorashi_2024,Li_2024,Chakraborty2024a,Debashish_2025}, which break time-reversal symmetry without net magnetization through momentum-dependent spin-splitting~\cite{ifmmode_2022,Libor_2022,Bhowal_2024,Mazin_2023}. Crucially, altermagnets enable field-free topological superconductivity, as demonstrated in d-wave altermagnet-proximitized Rashba nanowires~\cite{Ghorashi_2024,Debashish_2025}, making them ideal for robust Majorana platforms.

The superconducting diode effect (SDE) refers to direction-dependent critical current in superconducting devices~\cite{Daido_2022_intrinsic,nadeem_2023,Ma_2025,Nagaosa_2024}. Initially observed in low-symmetry superconductors~\cite{Broussard_1988,Jiang_1994,Papon_2008}, recent breakthroughs in engineered platforms, including superlattices~\cite{Narita_2024,ando_2020,sundaresh_2023}, bulk materials and thin films~\cite{Wakatsuki_2017,nadeem_2023,Yuki_2020,Schumann_2020}, multilayer graphene systems~\cite{lin_2022,Jaime_2023,Chen2025}, and transition metal dichalcogenides~\cite{bauriedl_2022,Yun_2023,Chen2025finite}, have renewed interest in its technological potential. The SDE requires simultaneous breaking of inversion and time-reversal symmetries~\cite{Wakatsuki_2017,Daido_2022_intrinsic,Nagaosa_2024}, typically achieved via external magnetic fields~\cite{ando_2020,sundaresh_2023,Hou_2023,gupta_2023,Abhishek_2023} or magnetic proximity~\cite{Yun_2023,narita_2022,gutfreund_2023} or in noncentrosymmetric superconductors with spontaneously broken time-reversal symmetry~\cite{Ghosh2020a,Shang2022Weyl,Shang2018,Shang2020,Sajilesh2025time}. A wide range of theoretical proposals for realizing SDE, using both phenomenological and microscopic approaches, have explored junction-based~\cite{Zhang_2022,Souto_2022,Steiner_2023,nadeem_2023} and junction-free configurations~\cite{nadeem_2023,Ma_2025,bhowmik_2025,Sayan_2024,Daido_2022_intrinsic,Yuan_2022,Ili_2022,Bhowmik2025PRB,He_2023,Picoli_2023}. Practical relevance is underscored by experimental demonstrations of high AC-to-DC conversion efficiencies~\cite{ingla_2025,castellani_2025}, affirming the promise of SDE for low-dissipation current control in quantum devices.

In junction-free systems, the superconducting diode effect is typically achieved by applying an external magnetic field to break time-reversal symmetry~\cite{nadeem_2023,Ma_2025}. However, relying on magnetic fields imposes severe constraints on scalable superconducting electronics: fields suppress the proximity-induced superconducting gap, generate detrimental flux noise, and are incompatible with densely integrated cryogenic circuits used in quantum technologies~\cite{nadeem_2023,Ma_2025,Nakamura2017,Rower2023}. To address this, we present a unified platform that intrinsically supports both field-free topological superconductivity and SDE in a junction-free architecture. Our setup consists of a one-dimensional (1D) spin spiral of magnetic adatoms deposited on a conventional three-dimensional (3D) $s$-wave superconductor, proximized by a $d$-wave altermagnet. This heterostructure intrinsically breaks both time-reversal and inversion symmetries that are essential prerequisites for the SDE, without relying on any applied magnetic field. Using a self-consistent Bogoliubov–de Gennes (BdG) mean-field approach, we demonstrate that the system supports a topological FF superconducting phase, tunable via the Cooper pair momentum controlled by an injected current. Within this FF regime, we show that the system exhibits highly efficient superconducting diode behavior, achieving efficiencies exceeding $\sim 45\%$ in the helical spin configuration and $\sim 35\%$ for the conical case. Our results offer a versatile and experimentally relevant platform that combines topological Majorana physics with diode functionality under entirely field-free conditions.

The remainder of the paper is organized as follows. In \sect{sec:real_space_hamiltonian}, we introduce the model Hamiltonian for the Shiba chain device proximitized by a $d$-wave altermagnet. Our main findings on the FF superconducting phase, including its topological properties and nonreciprocal charge transport, are presented in \sect{sec:result}. The topological features are discussed in \sect{subsec:topology}, followed by an analysis of the SDE for the helical and conical spin textures in \sect{subsec:sde}. We conclude in \sect{sec:summary} with a summary and experimental feasibility of our proposal, and an outlook for future works.

%%%%%%%%%%%%%%%%%%%%%%%%%%%%%%%%%%%%%%%%%%%%%%%%%%%%%%%%%%%%%%%%%%%%%%%%%%%%%%%%%%%%%%%%%%%%%%%%%%%%%%%%%%%%%%%%%%%%%%%%%%%%%%

\begin{figure}[!t]
\centering 
\includegraphics[width=\columnwidth]{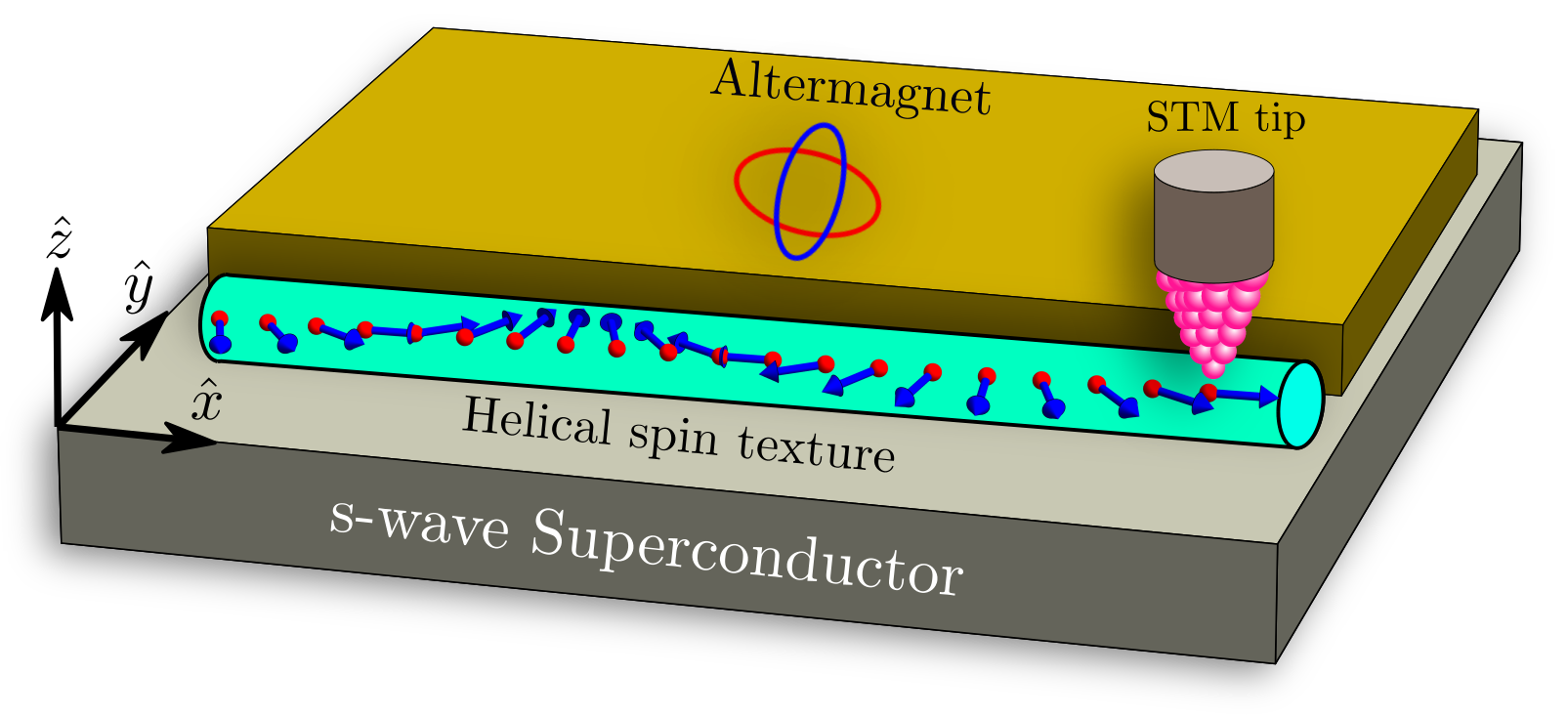}
\caption{\textbf{Schematic of an altermagnet-proximitized Shiba chain:} A 1D chain of magnetic adatoms with noncollinear classical spins is placed on the surface of a 3D $s$-wave superconductor and proximally coupled to a $d_{x^2-y^2}$-wave altermagnet.}   
\label{fig:model}
\end{figure}

%%%%%%%%%%%%%%%%%%%%%%%%%%%%%%%%%%%%%%%%%%%%%%%%%%%%%%%%%%%%%%%%%%%%%%%%%%%%%%%%%%%%%%%%%%%%%%%%%%%%%%%%%%%%%%%%%%%%%%%%%%%%%%%%

%--------------------------------------------------------
\section{Model Hamiltonian}\label{sec:real_space_hamiltonian}
%--------------------------------------------------------
We consider a 1D spin spiral formed by classical magnetic impurity atoms with spatially varying orientations, deposited on the surface of a 3D $s$-wave superconductor. This heterostructure is further brought into proximity with a $d_{x^2 - y^2}$ altermagnet which is known to support MZMs in 1D Rashba nanowires unlike its $d_{xy}$ counterpart~\cite{Ghorashi_2024}, as schematically illustrated in \fig{fig:model}. Since altermagnets carry zero net magnetization~\cite{Song2025altermagnets,Chen2025Electrical}, the induced altermagnetic exchange field does not reorient the classical impurity spins with large magnetic moments; hence, the spin-spiral texture is taken to remain unchanged. In realistic devices, interface imperfections can cause spatial variations in the proximity-induced altermagnetic strength, but to capture the essential physics we assume a smooth interface yielding a uniform coupling. Within this framework, we explore how the interplay between proximity-induced altermagnetism and conventional $s$-wave pairing in the bulk superconductor shapes the superconducting ground state of the Shiba chain. Under a supercurrent, this system can stabilize an FF state with finite-momentum pairing. To capture this physics, we model the parent $s$-wave superconductivity as realized by an attractive on-site Hubbard interaction:
\beq
\mathcal{H}_I = - \frac{U}{2} \sum_{s,s^\prime} \int \, d^3\mathbf{r}\,c^{\dagger}_s(\mathbf{r}) c^{\dagger}_{s^\prime}(\mathbf{r}) c_{s^\prime}(\mathbf{r}) c_{s}(\mathbf{r}),
\eeq
where $U > 0$ is the interaction strength and $c_s(\mathbf{r})$ annihilates an electron with spin $s$ at position $\mathbf{r}$. Within the mean-field approximation, we decouple $\mathcal{H}_I$ in the $s$-wave FF channel and assume that the resulting pairing correlations are induced in the 1D spin chain via the proximity effect~\cite{bhowmik_2025}. This allows us to model the chain as an effective superconducting system with a self-consistently determined pairing amplitude~\cite{SelfConsis} in the presence of the proximity induced altermagnetism. Based on this framework, we construct an effective BdG mean-field Hamiltonian for the Shiba chain device in \fig{fig:model} as:
\bea
    \mathcal{H} &=& \sum_{n=1}^{L} \Bigg[ \sum_{\sigma, \sigma^\prime} c_{n,\sigma}^\dagger \left\{ J \left( \mathbf{S}_n \cdot \boldsymbol{\sigma} \right)_{\sigma, \sigma^\prime} - \mu \delta_{\sigma,\sigma^\prime} \right\} c_{n,\sigma^\prime} \nonumber \\
    &+& \bigg\{ \sum_{\sigma, \sigma^\prime} c_{n,\sigma}^\dagger \left( \frac{J_A}{2} \left( \sigma_z \right)_{\sigma,\sigma^\prime} -t \delta_{\sigma, \sigma^\prime} \right) c_{n+1, \sigma^\prime} \nonumber \\
    &+& \Delta_n c_{n,\ua}^\dagger c_{n,\da}^\dagger + \textrm{h.~c.} \bigg\} \Bigg] + \epsilon_0 \;,\label{eq:BdG_ham_real} \\
    &=& \sum_{m} \left[E_m (q) \gamma_{m}^\dagger \gamma_m - \frac{E_m}{2} \right] + \epsilon_0 \;.\label{eq:hamiltonian}
\eea
\noindent Here, $\epsilon_0 = \frac{L}{U} |\Delta|^2$ is a constant energy shift, with $L$ denoting the total number of lattice sites. The parameters $t$, $J$, $J_A$ and $\mu$ correspond to the nearest-neighbor hopping amplitude, the exchange coupling between the magnetic adatoms and the itinerant electron spins, the proximity-induced altermagnetic strength and the chemical potential, respectively. The local spin vector of the impurity adatoms is denoted by $\mathbf{S}_n = \left\{\sin \theta \cos[\phi(n)],~\sin \theta \sin[\phi(n)],~\cos \theta \right\}$, realizing a spin-spiral texture along the chain with a uniform winding rate defined by $[\phi(n+1) - \phi(n)] = g$. Depending on the polar angle $\theta$, the spin configuration realizes different textures: for $\theta = \pi/2$, spins lie entirely in the $x$–$y$ plane, forming a helical spiral; for $0 < \theta < \pi/2$, the spins acquire a finite $z$-component and trace out a cone around the spiral axis, giving rise to a conical spiral. The $s$-wave FF order parameter is taken to be $\Delta_n = -i\sigma_y \Delta_0 e^{iqn}$, where $q$ denotes the Cooper pair momentum. $E_m(q)$ represents the Bogoliubov quasiparticle energies, and the operators $\gamma_m^\dagger$ ($\gamma_m$) create (annihilate) a Bogoliubov quasiparticle in the state $m$.

%%%%%%%%%%%%%%%%%%%%%%%%%%%%%%%%%%%%%%%%%%%%%%%%%%%%%%%%%%%%%%%%%%%%%%%%%%%%%%%%%%%%%%%%%%%%%%%%%%%%%%%%%%%%%%%%%%%%%%%%%%%%%%%%%%%%%%
%%%%%%%%%%%%%%%%%%%%%%%%%%%%%%%%%%%%%%%%%%%%%%%%%%%%%%%%%%%%%%%%%%%%%%%%%%%%%%%%%%%%%%%%%%%%%%%%%%%%%%%%%%%%%%%%%%%%%%%%%%%%%%%%%%%%%%

\begin{figure}[!t]
\centering 
\includegraphics[width=\columnwidth]{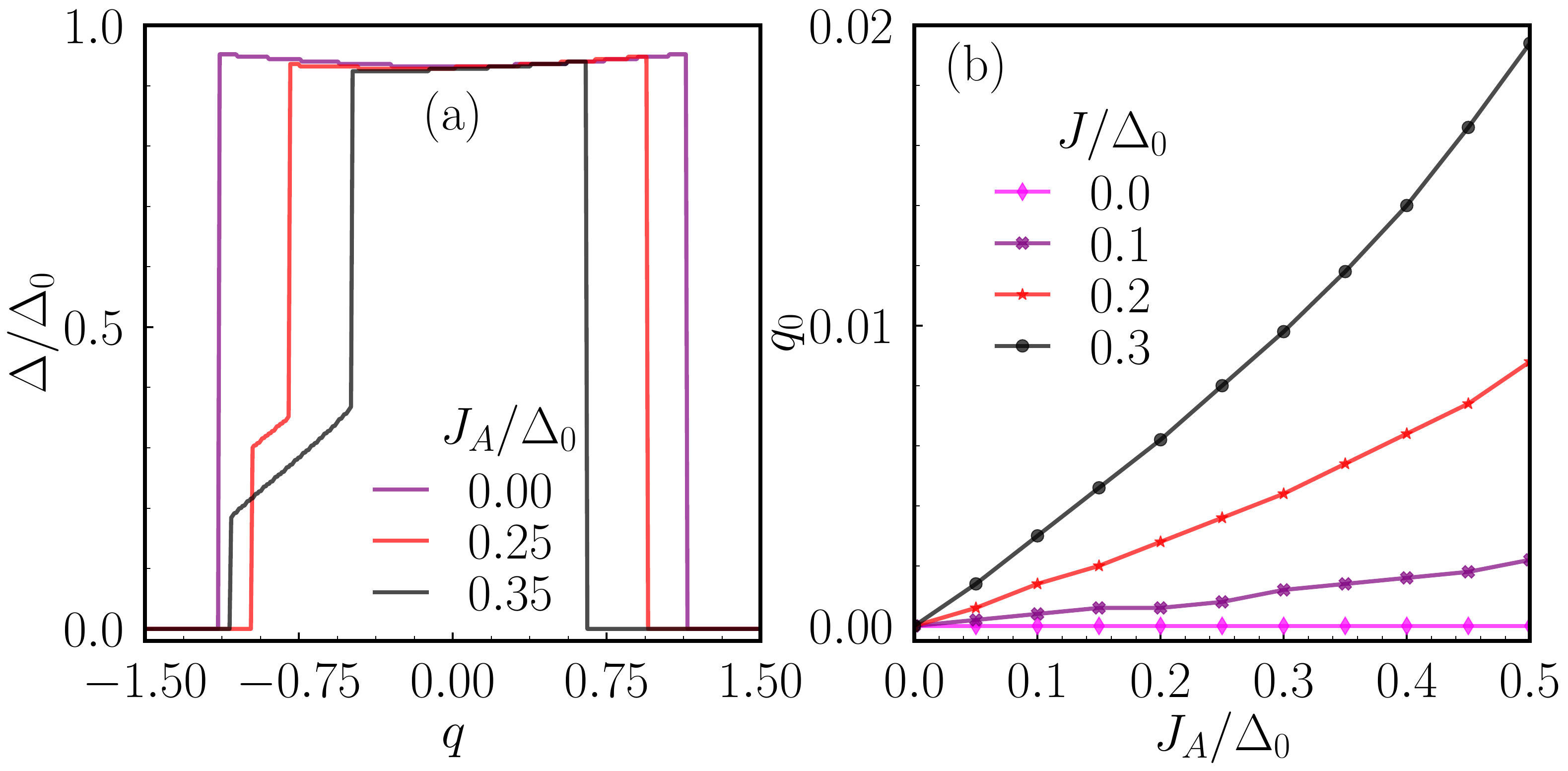}
\caption{\textbf{FF ground state for a helical Shiba chain:} (a) Self-consistent superconducting gap $\Delta(q)$ plotted as a function of the Cooper pair momentum $q$ for exchange coupling $J/\Delta_0 = 0.65$. (b) Optimal Cooper pair momentum $q_0$ corresponding to the FF ground state as a function of the proximity-induced altermagnetic field strength $J_A$.  The model parameters used are: ($t/\Delta_0$, $g$, $U/\Delta_0$, $\mu/\Delta_0$, $(\beta \Delta_0)^{-1}$) = ($0.5$, $\pi/2$, $1.38$, $1.0$, $0.01$).}   
\label{fig:fflo_real}
\end{figure}

%%%%%%%%%%%%%%%%%%%%%%%%%%%%%%%%%%%%%%%%%%%%%%%%%%%%%%%%%%%%%%%%%%%%%%%%%%%%%%%%%%%%%%%%%%%%%%%%%%%%%%%%%%%%%%%%%%%%%%%%%%%%%%%%%%%%%%
%%%%%%%%%%%%%%%%%%%%%%%%%%%%%%%%%%%%%%%%%%%%%%%%%%%%%%%%%%%%%%%%%%%%%%%%%%%%%%%%%%%%%%%%%%%%%%%%%%%%%%%%%%%%%%%%%%%%%%%%%%%%%%%%%%%%%%

The condensation energy $\Omega(q)$ of the system, defined as the difference of free energy per unit length between the superconducting and normal state, is
\begin{equation}
    \Omega(q, \Delta) = F(q, \Delta) - F(q, 0) \label{eq:cond_energy}\;.
\end{equation}

\noindent $F(q,\Delta)$ is the free energy density of the system given by
\begin{equation}
    F (q, \Delta) = - \frac{1}{L\beta} \sum_m \ln \left[ 1 + e^{-\beta E_m(q)} \right] + \frac{|\Delta|^2}{U}\,, \label{eq:free_energy}
\end{equation}
\noindent where, $\beta = (k_B T)^{-1}$ with $k_B$ being the Boltzmann constant and $T$ is the temperature. 

We determine the order parameter $\Delta (q)$ for a given value of $q$ by self-consistently solving the gap equation, obtained via minimizing the condensation energy $\Omega (q, \Delta)$ in \eqn{eq:cond_energy}, given by

\begin{equation}
    \Delta(q) = - \frac{U}{L} \sum_m \frac{\partial E_m}{\partial \Delta^*} n_F (E_m)\,. \label{eq:self_consistent}
\end{equation}

\noindent Here, $n_F(E_m) = 1/(1 + e^{\beta E_m})$ denotes the Fermi-Dirac distribution function. Using the self-consistent solution for $\Delta(q)$ in \eqn{eq:cond_energy}, we optimize the condensation energy with respect to $q$ to determine the Cooper pair momentum $q_0$ corresponding to the true FF superconducting ground state. The superconducting gap of the bulk 3D s-wave superconductor is denoted by $\Delta_0$, and the Hubbard interaction strength $U$ is chosen such that the system remains in the weak-coupling BCS regime with $\Delta_0 \sim 1$ meV.

The variation of the self-consistent superconducting order parameter $\Delta(q)$ and the Cooper pair momentum $q_0$ in the FF ground state of the helical spin chain as a function of the induced altermagnetic strength $J_A$ is shown in \fig{fig:fflo_real}, considering an in-plane spin texture ($\theta = \pi/2$). As seen in \fig{fig:fflo_real}(a), the superconducting gap vanishes beyond critical values of the Cooper pair momentum, either above a positive threshold $q = q_c^+$ or below a negative threshold $q = q_c^-$. \fig{fig:fflo_real}(b) illustrates that a nonzero $q_0$ requires a nonzero exchange energy $J$ and, $q_0$ varies linearly with $J_A$ for $J_A \ll \Delta_0$ at a fixed $J$. This feature plays a crucial role in the emergence of current-driven topology and a finite SDE in the system, as discussed in detail in later sections.

%--------------------------------------------------------
\subsection{Nanowire Limit of the Altermagnetic Shiba Chain}
\label{sec:momentum_space_hamiltonian}
%--------------------------------------------------------
To elucidate the underlying physics, we map the real-space altermagnetic Shiba-chain model in \eqn{eq:BdG_ham_real} onto an effective nanowire description, where the roles of spin–orbit coupling, exchange fields, and altermagnetic terms become transparent. Unlike the original Shiba-limit Hamiltonian, which lacks translational symmetry due to the spatially modulated spin texture, the nanowire formulation restores translation invariance, enabling a clear interpretation of the emergent FF pairing and band asymmetry. Starting from the low-energy continuum limit of \eqn{eq:BdG_ham_real}, we perform a local spin-dependent gauge transformation~\cite{Hess_2022,Pritam2024} followed by a Fourier transformation, yielding the continuum Hamiltonian:
\bea
    \tilde{h}_k &=& \xi_{k,\tilde{g}} - \mu - \frac{\hbar^2}{m} \tilde{g} k \sigma_z + J \sin \theta \sigma_x + J \cos \theta \sigma_z \non \\
    &+& J_A \bigg(1 - \frac{1}{2} (k^2 + \tilde{g}^2) \bigg) \sigma_z + J_A \tilde{g} k\,,
\eea
\noin where $\xi_{k,\tilde{g}} = \frac{\hbar^2}{2m} (k^2 + \tilde{g}^2)$, with $\tilde{g} = g/2$. Here, the term proportional to $\tilde{g}k\sigma_z$ represents an emergent spin–orbit coupling originating from the spin texture, while the altermagnetic contributions modify the kinetic energy in a manner equivalent to coupling to an effective gauge field. To proceed, we lattice-regularize the continuum Hamiltonian using the substitutions $k \rightarrow \sin k$ and $1 - \frac{k^2}{2} \rightarrow \cos k$~\cite{Pritam2024}. This yields an effective mean-field BdG Hamiltonian (see the Supplementary Material (SM) for details) describing the s-wave FF superconducting state in the basis $\psi(k,q)=\left( c_{k+ q/2,\ua}, c_{k+ q/2,\da}, c_{-k+ q/2,\da}^\dagger, -c_{-k+ q/2,\ua}^\dagger \right)^T$:
\bea
\mathcal{H}^\prime &=& \frac{1}{2} \sum_{k} \psi^\dagger(k,q) H_k (q) \psi(k,q) + \epsilon_0 \,, \non\\
&=& \sum_{n,k} \left(E_{n,k} \gamma^\dagger_{n,k} \gamma_{n,k} - \frac{E_{n,k}}{2} \right) + \epsilon_0 \,, \non\\
H_k(q) &=&
    \begin{bmatrix}
	h_{k+q/2} & -i \sigma_y \Delta \\
	i \sigma_y \Delta & -h^{*}_{-k + q/2}
    \end{bmatrix}\,, \non
\eea
\vspace{-0.3cm}
\bea
    \text{\rm with,}\;\;h_k &=& \eps_{k,\tilde{g}} - t~g \sin k \, \sigma_z + J \sin \theta \sigma_x + J \cos \theta \sigma_z \non \\
    &+& J_A \bigg(\cos k - \frac{\tilde{g}^2}{2}\bigg) \sigma_z + J_A \tilde{g} \sin k\,. \label{eq:k_space_ham}
\eea
\noindent Here, $\eps_{k,\tilde{g}} = t~(2 - 2\cos k) + t~\tilde{g}^2 - \mu$ with $t=\frac{\hbar^2}{2m}$ and we set $\hbar = m = 1$ throughout our analysis for simplicity. The bogoliubov quasiparticle spectrum exhibits distinct symmetry-breaking behavior depending on the type of spin texture: for a helical chain, spectral asymmetry arises only when both $J$ and $J_A$ are finite, while for a conical chain, $J$ alone suffices due to intrinsic inversion symmetry breaking from the spin texture (See the SM for details). These features underlie the emergence of superconducting diode response in our system.\\

%%%%%%%%%%%%%%%%%%%%%%%%%%%%%%%%%%%%%%%%%%%%%%%%%%%%%%%%%%%%%%%%%%%%%%%%%%%%%%%%%%%%%%%%%%%%%%%%%%%%%%%%%%%%%%%%%%%%%%%%%%%%%%%%%%%%%%%%%%

\begin{figure}[!t]
\centering 
\includegraphics[width=\columnwidth]{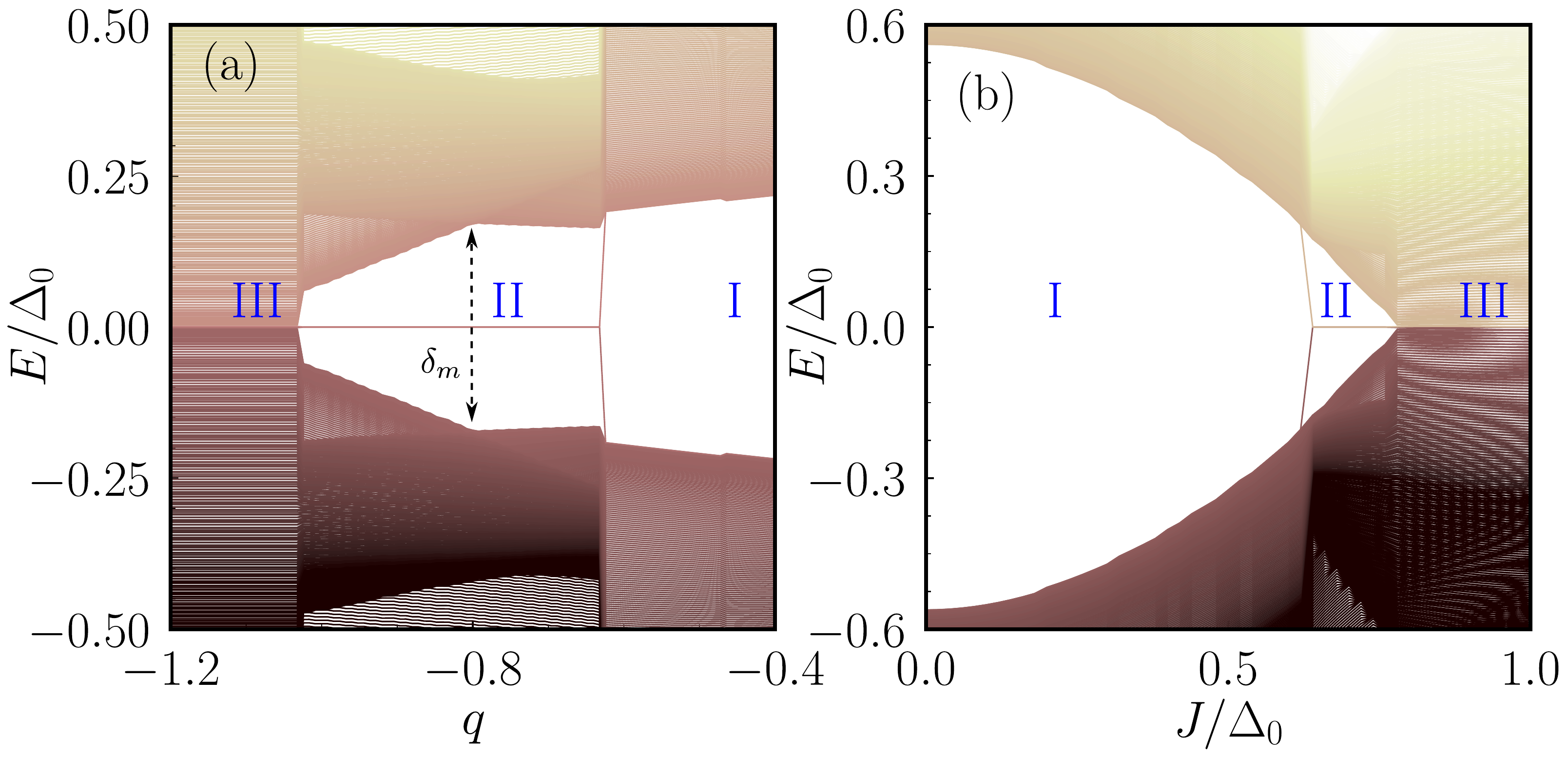}
\caption{\textbf{Topological phase transition and the emergence of MZMs:} (a) Bogoliubov quasiparicle energy spectrum under open boundary conditions as a function of Cooper pair momentum $q$, incorporating the self-consistently calculated pairing gap $\Delta(q)$ for exchange coupling $J/\Delta_0 = 0.65$ and altermagnetic exchange strength $J_A/\Delta_0 = 0.3$. (b) Spectrum plotted as a function of $J$ for fixed $q = -0.8$ and $J_A/\Delta_0 = 0.3$. The system exhibits three distinct phases: I) a trivial superconducting phase, II) a topological superconducting phase hosting MZMs, protected by a finite minigap $\delta_m$, and III) a normal phase where the superconducting order vanishes. Calculations are performed for a chain of 400 sites with model parameters: ($t/\Delta_0$, $g$, $U/\Delta_0$, $\mu/\Delta_0$, $(\beta~\Delta_0)^{-1}$) = ($0.5$, $\pi/2$, $1.38$, $1.0$, $0.01$).}   
\label{fig:spectrum_obc}
\end{figure}

%%%%%%%%%%%%%%%%%%%%%%%%%%%%%%%%%%%%%%%%%%%%%%%%%%%%%%%%%%%%%%%%%%%%%%%%%%%%%%%%%%%%%%%%%%%%%%%%%%%%%%%%%%%%%%%%%%%%%%%%%%%%%%%%%%%%%%%%%%

%---------------------------------------------------------
\section{Results and discussions}\label{sec:result}
%---------------------------------------------------------
In this section, we present our main results in two parts. First, we analyze the emergence of topology in the FF superconducting state in the Shiba chain proximitized by a $d$-wave altermagnet. Then, we examine the field-free SDE arising from the interplay between magnetic spin textures and the induced altermagnetism, focusing on helical and conical configurations. These textures produce distinct diode responses, underscoring the key role of magnetic structure in nonreciprocal transport.

%%%%%%%%%%%%%%%%%%%%%%%%%%%%%%%%%%%%%%%%%%%%%%%%%%%%%%%%%%%%%%%%%%%%%%%%%%%%%%%%%%%%%%%%%%%%%%%%%%%%%%%%%%%%%%%%%%%%%%%%%%%%%%%%%%%%%%%%%%

\begin{figure}[!t]
\centering 
\includegraphics[width=\columnwidth]{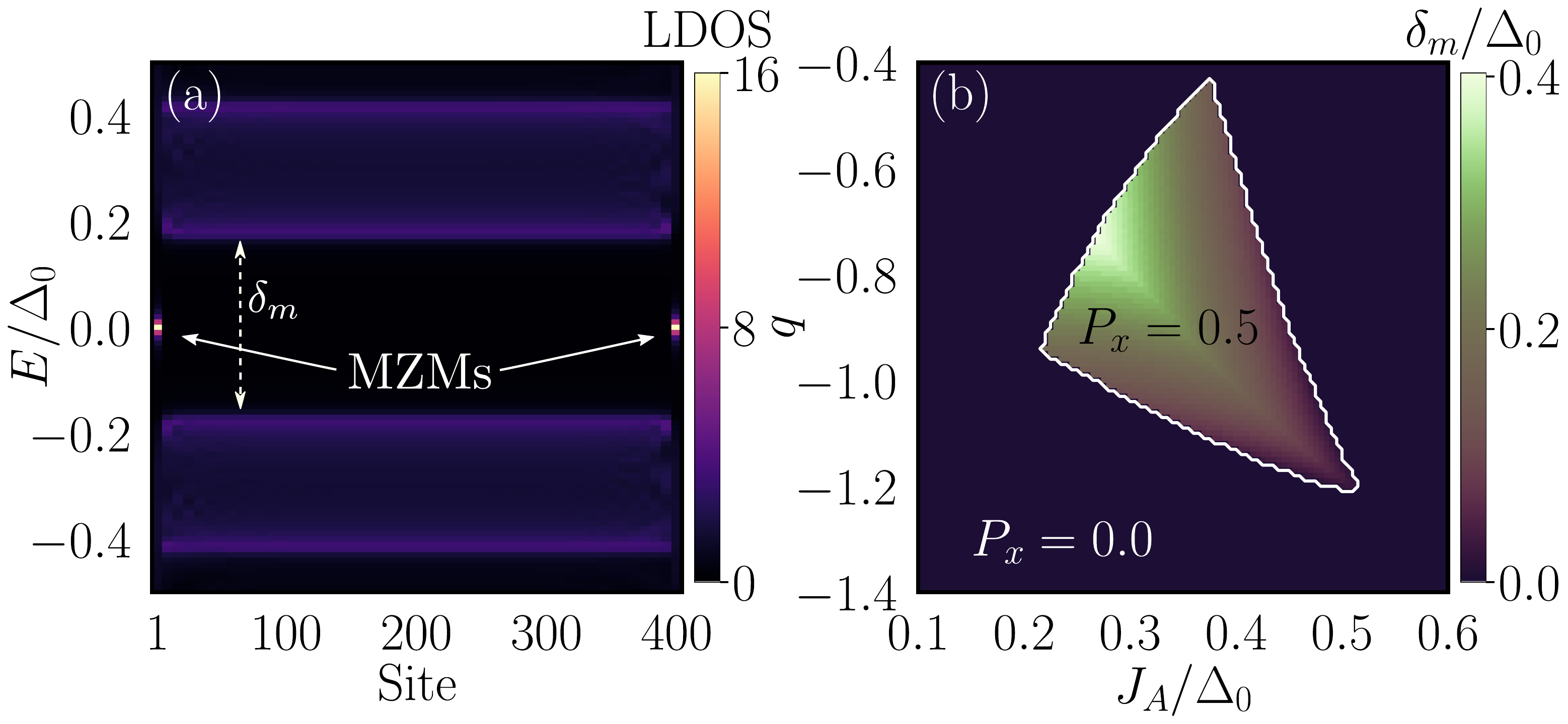}
\caption{\textbf{Topological signatures and local density of states (LDOS) of the MZMs:} (a) Site– and energy–resolved LDOS profile clearly demonstrates strong localization of MZMs at the ends of the chain for fixed $q = -0.8$ and $J_A/\Delta_0 = 0.3$. The LDOS also shows the Shiba bands inside the superconducting gap separated by a minigap $(\delta_m)$. (b) Topological phase diagram showing the minigap $\delta_m$ as functions of $J_A$ and $q$. The white contour delineates the phase boundary separating the topological regime $(P_x=0.5)$ from trivial regime $(P_x=0)$, indicating that a finite minigap protects the MZMs. Other Model parameters are: ($t/\Delta_0$, $g$, $U/\Delta_0$, $\mu/\Delta_0$, $(\beta \Delta_0)^{-1}$, $J/\Delta_0$) = ($0.5$, $\pi/2$, $1.38$, $1.0$, $0.01$, $0.65$).}   
\label{fig:topology}
\end{figure}

%%%%%%%%%%%%%%%%%%%%%%%%%%%%%%%%%%%%%%%%%%%%%%%%%%%%%%%%%%%%%%%%%%%%%%%%%%%%%%%%%%%%%%%%%%%%%%%%%%%%%%%%%%%%%%%%%%%%%%%%%%%%%%%%%%%%%%%%%%

%%%%%%%%%%%%%%%%%%%%%%%%%%%%%%%%%%%%%%%%%%%%%%%%%%%%%%%%%%%%%%%%%%%%%%%%%%%%%%%%%%%%%%%%%%%%%%%%%%%%%%%%%%%%%%%

\begin{figure*}[!htb]
\centering 
\includegraphics[width=\linewidth]{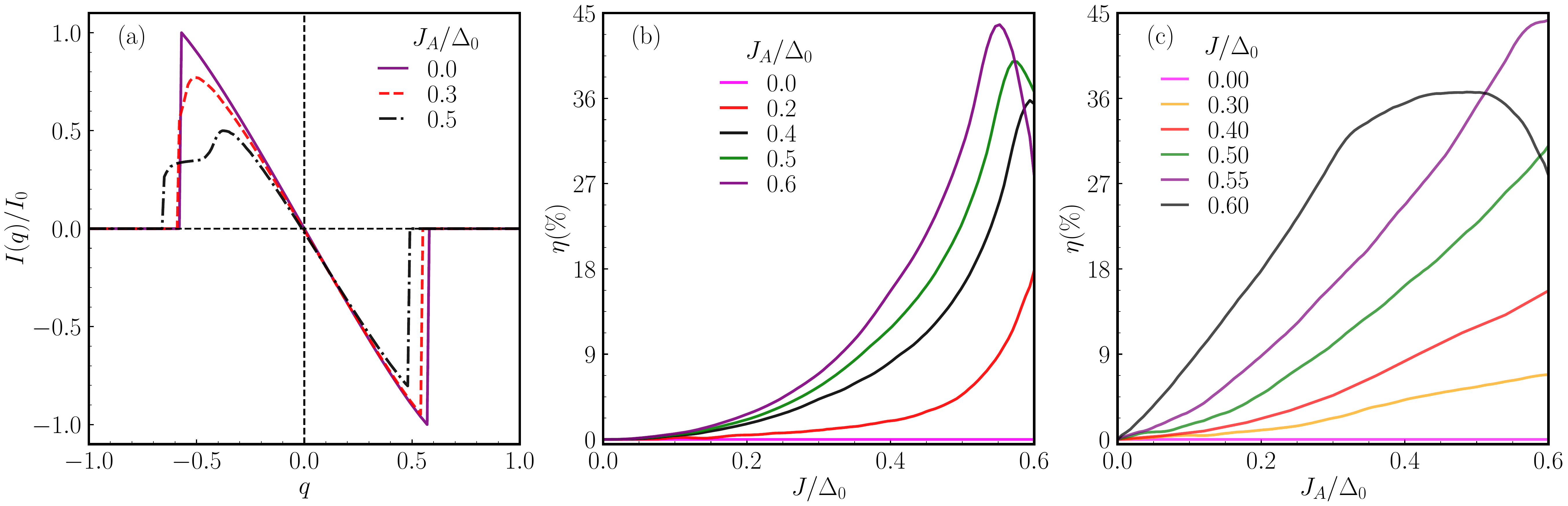}
\caption{\textbf{Field-free SDE with a helical spin texture:} (a) Supercurrent density $I(q)$ as a function of Cooper pair momentum $q$ for $J/\Delta_0 = 0.5$. Nonreciprocal behavior ($I(q) \ne -I(-q)$) arises only when both the magnetic exchange $J$ and the altermagnetic coupling $J_A$ are finite, leading to an asymmetry in critical currents $|I_c^+| \ne |I_c^-|$ and a nonzero diode efficiency $\eta \ne 0$. Here, $I_0 \equiv |I_c^+(J_A = 0)| = |I_c^-(J_A = 0)|$. (b) and (c): Diode efficiency $\eta$ as a function of $J$ (at fixed $J_A$) and $J_A$ (at fixed $J$), respectively. Model parameters: ($t/\Delta_0$, $g$, $\mu/\Delta_0$, $(\beta~\Delta_0)^{-1}$, $U/\Delta_0$) = ($1.0$, $2\pi/3$, $1.0$, $0.1$, $1.574$).}   
\label{fig:sde_helical_real}
\end{figure*}

%%%%%%%%%%%%%%%%%%%%%%%%%%%%%%%%%%%%%%%%%%%%%%%%%%%%%%%%%%%%%%%%%%%%%%%%%%%%%%%%%%%%%%%%%%%%%%%%%%%%%%%%%%%%%%%

%---------------------------------------------------------
\subsection{Topologically protected MZMs in FF superconducting phase}\label{subsec:topology}
%---------------------------------------------------------
We investigate the topological properties of the helical Shiba chain proximitized to a $d$-wave altermagnet by analyzing the BdG Hamiltonian in \eqn{eq:hamiltonian} under open boundary conditions. Using the self-consistently determined pairing amplitude $\Delta$, we compute the quasiparticle energy spectrum as a function of the Cooper pair momentum ($q$) and the exchange coupling ($J$), as shown in \fig{fig:spectrum_obc}. The resulting spectra shown in \fig{fig:spectrum_obc}(a) and \fig{fig:spectrum_obc}(b) reveal three distinct regimes: I) a trivially gapped superconducting phase, II) a topological superconducting phase characterized by doubly degenerate zero-energy in-gap modes, signaling the presence of MZMs stabilized by a finite supercurrent, and III) a gapless normal metallic phase where superconductivity is completely suppressed ($\Delta = 0$). Notably, \fig{fig:spectrum_obc}(a) shows that the topological phase boundaries shift with the Cooper pair momentum $q$, demonstrating that the supercurrent itself provides an experimentally tunable handle for accessing and controlling the topological superconducting phase~\cite{Takasan2022}.

We further examine the MZMs by computing the site-resolved local density of states (LDOS), shown in \fig{fig:topology}(a). In the topological regime, the LDOS at the chain ends exhibits a sharp zero-energy peak, confirming the presence of topologically protected MZMs separated from bulk quasiparticle excitations by a finite minigap~$\delta_m$. In contrast, the LDOS in the middle of the chain displays only subgap Yu–Shiba–Rusinov (YSR) bands residing within the induced superconducting gap, as illustrated in the inset of \fig{fig:topology}(a).\\

\noindent\textit{\textbf{Topological characterization:}} For a quantitative topological characterization of the regime-II, we employ the bulk polarization ($P_x$)~\cite{Resta_1998,Wheeler_2019,Kang_2019} as a diagnostic topological invariant, allowing us to distinguish the topological regime from the trivial one. It is defined as~\cite{Resta_1998,Wheeler_2019,Kang_2019}:

\begin{equation}
    P_x = \frac{1}{2\pi} \mathrm{Im}\left[ \mathrm{Tr}\left\{ \ln\left( \mathcal{U}^\dagger \mathcal{W} \mathcal{U} \right) \right\} \right]\,,
\end{equation}

\noindent where, $\mathcal{U}$ is an $N \times N_{\text{occ}}$ matrix constructed by columnwise stacking the occupied energy eigenstates of the BdG lattice Hamiltonian in \eqn{eq:hamiltonian}. The operator $\mathcal{W} = \exp\left(i 2\pi \hat{p}\right)$ is built from the microscopic dipole operator $\hat{p} = \hat{x}
/L$, with $L$ denoting the system size and $\hat{x}$ the position operator. A value of $P_x = 0.5$ identifies the topological phase hosting MZMs, whereas $P_x = 0$ indicates the trivially gapped phase~\cite{Resta_1998,Wheeler_2019,Kang_2019}. We have calculated the bulk polarization ($P_x$) and the minigap ($\delta_m$) within the $J_A$–$q$ plane, using the self-consistently obtained $\Delta$, as shown in \fig{fig:topology}(b). \fig{fig:topology}(b) reveals the topological FF phase characterized by a bulk polarization value of $P_x = 0.5$. The minigap $\delta_m$, which separates the MZMs from the bulk superconducting states, serves as a crucial indicator of topological protection. The finite $\delta_m$ in the topological regime observed in \fig{fig:topology}(b) confirms that the MZMs are robustly protected by the minigap. An analogous topological FF phase appears in the conical Shiba chain as well, characterized by $P_x = 0.5$ and a finite $\delta_m$. However, unlike the helical case, the conical Shiba chain supports the topological phase even in the absence of altermagnetic proximity (see SM for details).

The topological FF state realized in our system constitutes a distinct class of superconducting phases in which finite-momentum pairing coexists with nontrivial band topology. The nonzero pairing momentum fundamentally reshapes the quasiparticle band structure and redistributes Berry curvature, thereby enabling gapped topological phases that are inaccessible in conventional $q=0$ superconductors~\cite{Zhang2013topological, Qu2013topological}. These phases support robust Majorana zero modes and, in higher dimensions, chiral edge states. Importantly, the topological FF phase is highly tunable: its existence and stability depend sensitively on parameters such as spin–orbit coupling strength, chemical potential, and the direction or magnitude of effective Zeeman fields—or, as in our case, the externally controlled supercurrent that sets the pairing momentum. This enhanced tunability provides additional control knobs for driving topological phase transitions compared to standard topological superconductors. Topological FF phases have been predicted and explored in experimentally accessible platforms including spin–orbit–coupled cold atomic gases and semiconductor–superconductor heterostructures~\cite{Zhang2013topological, Qu2013topological}, offering promising routes toward realizing new Majorana-based quantum devices.

%---------------------------------------------------------
\subsection{Field-free superconducting diode effect}\label{subsec:sde}
%---------------------------------------------------------
Having established the topological nature of the FF state, we now examine the SDE-- a signature of finite-momentum pairing. In superconducting systems lacking both inversion and time-reversal symmetry, the critical current becomes nonreciprocal: the maximum dissipationless current differs along and against the $q_0$ direction. In our setup, this symmetry breaking stems from the interplay between the spin spiral and the $d$-wave altermagnet. To quantify the SDE, we compute the supercurrent density $I(q)$ defined as~\cite{Daido_2022_intrinsic,Nagaosa_2024,bhowmik_2025}:

\begin{equation}
    I(q)= - 2e \frac{\partial \Omega(q, \Delta)}{\partial q}\,. 
    \label{eq:current}
\end{equation}

\noindent The sign of $I(q)$ determines the direction of supercurrent flow. The maximum sustainable supercurrent in the positive and negative directions within the stable superconducting region $q_c^- \leq q \leq q_c^+$ are denoted by $I_c^+$ and $I_c^-$, respectively. A finite SDE emerges when $|I_c^+| \neq |I_c^-|$. The efficiency, or quality factor, of the SDE is then defined as:
\begin{equation}
    \eta = \frac{|I_c^-| - |I_c^+|}{|I_c^+| + |I_c^-|}\,.  \label{eq:efficiency}
\end{equation}
\noindent We first calculate $I(q)$ across the range $q_c^- \leq q \leq q_c^+$ using the self-consistent solutions for $\Delta(q)$ at each $q$ value. By identifying the critical supercurrent values $I_c^+$ and $I_c^-$, we numerically determine the diode efficiency ($\eta$). To thoroughly investigate the SDE, we systematically explore two distinct scenarios in our Shiba chain setup: (i) with in-plane helical spin texture (for $\theta=\pi/2$) and (ii) with out-of-plane or conical spin texture (for $0 < \theta < \pi / 2$).

%------------------------------------------------------------------
\subsubsection{Case-A: In-plane helical spin texture}
%------------------------------------------------------------------
The behavior of the supercurrent density $I(q)$ and the diode efficiency $\eta$ for the Shiba chain with helical spin texture under different system parameters are shown in \fig{fig:sde_helical_real}. \fig{fig:sde_helical_real}(a) demonstrates that $I(q)$ remains symmetric with respect to $q$ when the induced altermagnetic strength $J_A = 0$, even if the in-plane exchange field $J \ne 0$. This is because the helical spin texture preserves inversion symmetry up to a global spin rotation. However, as the magnitude of $J_A$ increases with finite $J$, the supercurrent becomes more and more asymmetric, resulting in nonreciprocal critical currents ($|I_c^+| \ne |I_c^-|$), and thereby a finite diode efficiency, indicating the onset of SDE. This behavior stems from the presence of an FF state and the resulting asymmetry in the critical momenta, $|q_c^+| \ne |q_c^-|$, which becomes evident only when both $J$ and $J_A$ are nonzero, as highlighted in \fig{fig:fflo_real}(a). \fig{fig:sde_helical_real}(b) and \fig{fig:sde_helical_real}(c) present $\eta$ as a function of $J$ for fixed values of $J_A$ and of $J_A$ for fixed values of $J$, respectively. In \fig{fig:sde_helical_real}(b), $\eta$ increases monotonically with $J$ and exhibits a nonlinear enhancement beyond $J/\Delta_0 \approx 0.3$, reaching a peak at intermediate values before declining. This trend reflects the competition between band asymmetry and change in the superconducting pairing susceptibility as the exchange field strength is varied, consistent with previous studies~\cite{Yuan_2022}. The maximum efficiency increases as one increases the magnitude of $J_A$. Likewise, \fig{fig:sde_helical_real}(c) shows that $\eta$ grows steadily with $J_A$, and the maximum efficiency improving for larger values of $J/\Delta_0$. Momentum-space analysis presented in the Supplementary Material shows that, by tuning $J$ and $J_A$, the helical Shiba chain can achieve a diode efficiency as high as $40 \%$ (see SM for details).

\begin{figure}[!t]
\centering 
\includegraphics[width=\columnwidth]{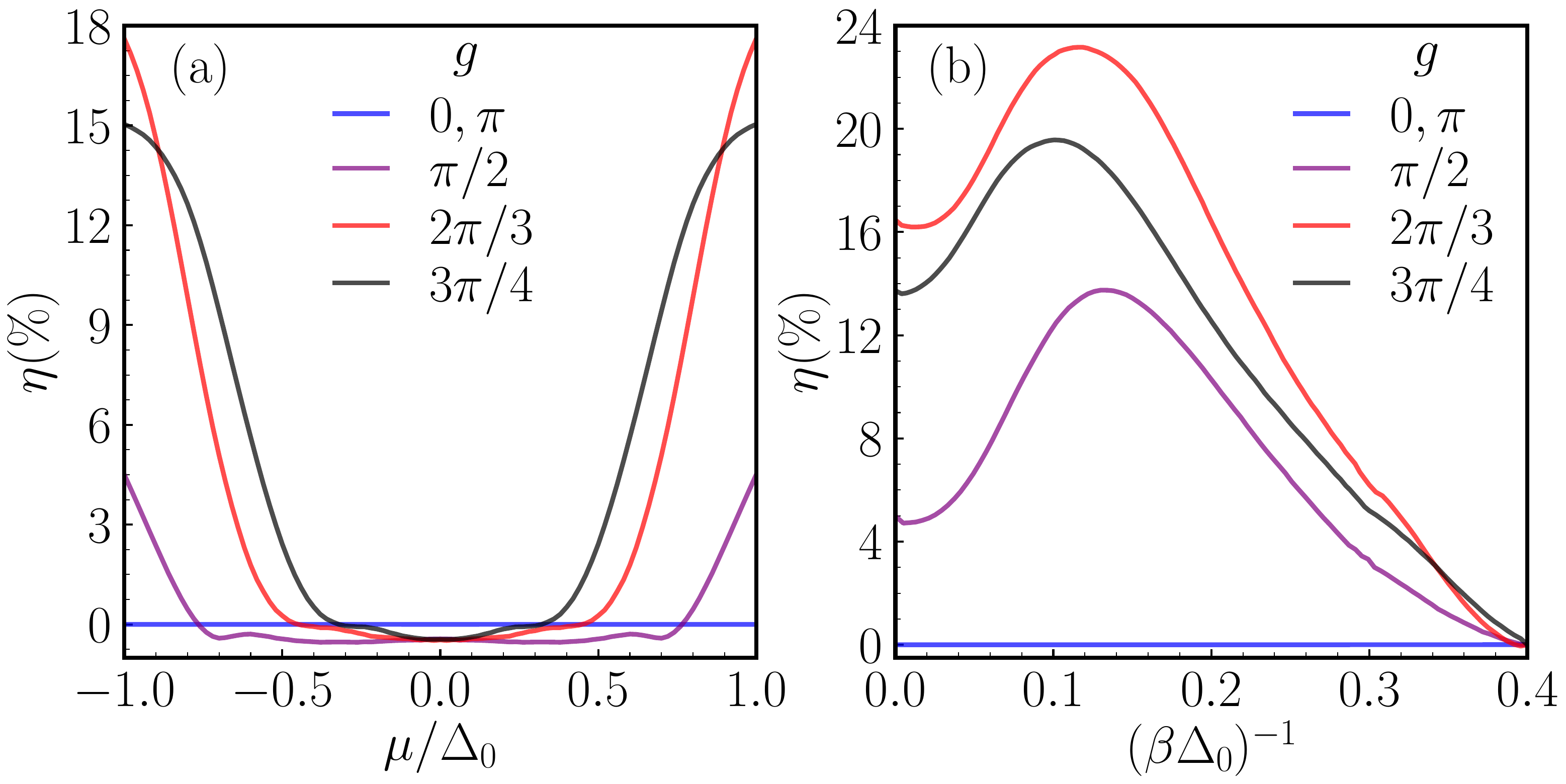}
\caption{\textbf{Diode efficiency for different types of helical spin texture:} (a) Variation of $\eta$ as a function of chemical potential $\mu$, calculated at a fixed temperature $(\beta \Delta_0)^{-1} = 0.01$ and interaction strength $U/\Delta_0 = 1.555$. (b) Temperature dependence of $\eta$ for fixed $\mu/\Delta_0 = 1.0$ and $U/\Delta_0 = 1.574$. Common model parameters are: ($t/\Delta_0$, $J/\Delta_0$, $J_A/\Delta_0$) = ($1.0$, $0.5$, $0.5$).}   
\label{fig:efficiency_vs_mu_T}
\end{figure}

The dependence of SDE efficiency on the chemical potential ($\mu$) and temperature ($\beta^{-1}$) is shown in \fig{fig:efficiency_vs_mu_T} for various values of the angle $g$ between adjacent spins. The system does not exhibit any SDE for ferromagnetic ($g = 0$) or antiferromagnetic ($g = \pi$) spin configurations, as inversion symmetry remains preserved in both the cases. In contrast, a spin spiral configuration ($0 < g < \pi$) breaks both inversion and time-reversal symmetries, giving rise to a finite SDE. As the chemical potential increases from zero, the diode efficiency $\eta$ initially decreases and vanishes at a critical value. Beyond this point, further increase in $\mu$ leads to a sign reversal of $\eta$, which then grows in magnitude and reaches a maximum near $\mu = 1.0$ [\fig{fig:efficiency_vs_mu_T}(a)]. Notably, the efficiency curve is symmetric about $\mu = 0$, with identical behavior for positive and negative values of the chemical potential. In addition, the temperature dependence of $\eta$ is shown in \fig{fig:efficiency_vs_mu_T}(b), which reveals that the efficiency initially increases with temperature and reaches a maximum around $(\beta~\Delta_0)^{-1} = 0.1$. Beyond this point, further increase in temperature suppresses the diode response, and the efficiency eventually vanishes.

%%%%%%%%%%%%%%%%%%%%%%%%%%%%%%%%%%%%%%%%%%%%%%%%%%%%%%%%%%%%%%%%%%%%%%%%%%%%%%%%%%%%%%%%%%%%%%%%%%%%%%%%%%%%%%%%%%%%%%%%%%%%%%%%%%%%%%%%%%

%%%%%%%%%%%%%%%%%%%%%%%%%%%%%%%%%%%%%%%%%%%%%%%%%%%%%%%%%%%%%%%%%%%%%%%%%%%%%%%%%%%%%%%%%%%%%%%%%%%%%%%%%%%%%%%%%%%%%%%%%%%%%%%%%%%%%%%%%%

%%%%%%%%%%%%%%%%%%%%%%%%%%%%%%%%%%%%%%%%%%%%%%%%%%%%%%%%%%%%%%%%%%%%%%%%%%%%%%%%%%%%%%%%%%%%%%%%%%%%%%%%%%%%%%%

\begin{figure*}[!htb]
\centering 
\includegraphics[width=\linewidth]{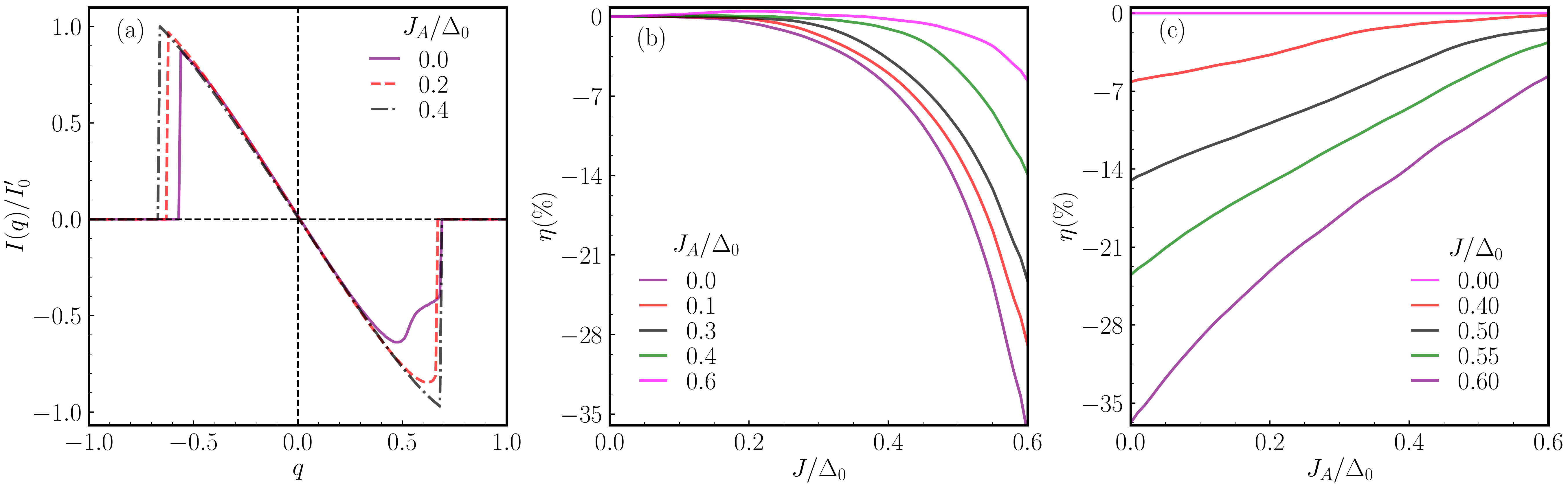}
\caption{\textbf{Field-free SDE with a conical spin texture:} (a) The supercurrent density ($I$) is shown as a function of $q$, considering $J/\Delta_0 = 0.5$.  Nonreciprocal behavior, indicated by $I(q) \ne -I(-q)$, emerges even with $J_A=0$. This leads to an asymmetry in the critical supercurrents, $|I_c^+| \ne |I_c^-|$, resulting in a finite diode efficiency $\eta \ne 0$. Here, $I_{0}^\prime \equiv |I_c^+(J_A/\Delta_0 = 0.4)| = |I_c^-(J_A/\Delta_0 = 0.4)|$. (b) and (c): The corresponding diode efficiency ($\eta$) is shown as a function of $J$ (with fixed $J_A$) and $J_A$ (with fixed $J$), respectively. Other model parameters are ($t/\Delta_0$, $g$, $\mu/\Delta_0$, $(\beta~\Delta_0)^{-1}$, $U/\Delta_0$) = ($1.0$, $2 \pi/3$, $1.0$, $0.01$, $1.574$).}   
\label{fig:sde_conical_real}
\end{figure*}

%%%%%%%%%%%%%%%%%%%%%%%%%%%%%%%%%%%%%%%%%%%%%%%%%%%%%%%%%%%%%%%%%%%%%%%%%%%%%%%%%%%%%%%%%%%%%%%%%%%%%%%%%%%%%%%

%-----------------------------------------------------------------
\subsubsection{Case-B: Out of plane conical spin texture}
%------------------------------------------------------------------
We analyze $I(q)$ and $\eta$ for a Shiba chain with a conical spin texture, shown in \fig{fig:sde_conical_real}. As seen in \fig{fig:sde_conical_real}(a), the supercurrent is asymmetric even when $J_A = 0$, provided $J$ is finite. This contrasts with the helical case, where both $J$ and $J_A$ are needed to break inversion and time-reversal symmetries. In the conical texture, these symmetries are inherently broken by the out-of-plane spin component, enabling nonreciprocal current flow even without altermagnetism. Interestingly, increasing the magnitude of $J_A$ while keeping $J$ finite progressively suppresses this asymmetry, leading to a reduction in $\eta$. \fig{fig:sde_conical_real}(b) and \fig{fig:sde_conical_real}(c) further illustrate the variation of $\eta$ as a function of $J$ for fixed $J_A$ and of $J_A$ for fixed $J$, respectively. From \fig{fig:sde_conical_real}(b), we observe that $\eta$ increases nonlinearly with $J$, becoming prominent beyond $J/\Delta_0 \sim 0.2$. Notably, in contrast to the helical case where the efficiency is positive, the diode efficiency for the conical texture is negative. Moreover, the peak value of $\eta$ decreases progressively as $J_A$ increases. In contrast, \fig{fig:sde_conical_real}(c) shows that increasing $J_A$ leads to a gradual suppression of $\eta$, highlighting that the altermagnetic exchange competes with the intrinsic asymmetry generated by the conical spin texture and tends to restore symmetry in the quasiparticle dispersion.\\

\noindent A major challenge in realizing topological superconductivity in magnet–superconductor heterostructures~\cite{Rachel2025,Choi2019} is the reliance on external magnetic fields to break time-reversal symmetry, which often suppresses the proximity-induced superconducting gap. Recent theoretical advances have identified altermagnet–superconductor heterostructures as promising alternatives~\cite{Ghorashi_2024,Li_2024,Debashish_2025}, since they intrinsically break time-reversal symmetry while maintaining zero net magnetization. In our device, the $d$-wave altermagnet plays two roles. First, it induces the requisite time-reversal symmetry breaking to realize topological superconductivity in the Shiba chain without external fields, enabling the emergence of MZMs while preserving the superconducting gap. Second, in the presence of a helical spin texture in the chain, the altermagnet also breaks inversion symmetry via its momentum-dependent spin splitting. This combined symmetry breaking facilitates nonreciprocal supercurrent flow, resulting in a field-free SDE.

%---------------------------------------------------------
\section{Summary and conclusions}\label{sec:summary}
% %~~~~~~~~~~~~~~~~~~~~~~~~~~~~~~~~~~~~~
We have demonstrated that the altermagnet-proximitized Shiba chain device serves as a unified platform for realizing both field-free topological FF superconductivity and the SDE. Through self-consistent real-space BdG calculations, we showed the emergence of an asymmetric FF state that supports two key functionalities: (i) a topological superconducting phase characterized by a bulk polarization $P_x = 0.5$ and protected by a finite minigap ($\delta_m$), hosting MZMs at the chain ends tunable via an externally injected supercurrent; and (ii) a strong nonreciprocal supercurrent response $(|I_{c}^+| \ne |I_{c}^-|)$, yielding diode efficiencies of $\eta \gtrsim 45\%$ for helical and $\eta \gtrsim 35\%$ for conical spin textures—all achieved without any applied magnetic field. Because the Cooper-pair momentum ($q$) is set by the injected supercurrent, the phase boundaries of the topological FF states shift with $q$, providing a direct and experimentally tunable handle for accessing and controlling the topological regime. This finite-momentum pairing also modifies the quasiparticle spectrum in ways not possible at $q=0$, enabling gapped topological phases and Majorana modes previously identified in topological FF states in cold atomic platforms~\cite{Zhang2013topological,Qu2013topological}. The symmetry-breaking mechanisms underlying these effects are qualitatively distinct: in helical Shiba chains, both time-reversal and inversion symmetries are broken via proximity to a $d$-wave altermagnet due to the orthogonal orientation of spin–orbit and exchange fields. In contrast, conical Shiba chains intrinsically break inversion symmetry via their out-of-plane spin components, resulting in spectral asymmetry and nonreciprocal transport. The $d$-wave altermagnet thus plays a dual and indispensable role, offering a symmetry-engineered pathway to simultaneously achieve topological superconductivity and superconducting diode functionality in a junction-free, magnetic field-free architecture.

Our proposal is well-aligned with current experimental capabilities in atomic-scale engineering of magnet–superconductor heterostructures~\cite{Rachel2025,Choi2019,lo2024magnet}. Shiba chains assembled from Fe, Co or Mn adatoms on conventional $s$-wave superconductors such as Pb(110), Nb(110) or Re(0001) have been shown to support both helical and conical spin textures~\cite{Stevan_2014,schneider_2022,Kim_2018,Ruby_2015}. These magnetic structures can be proximitized with $d$-wave altermagnetic materials like RuO$_2$~\cite{ifmmode_2022}, MnF$_2$~\cite{Bhowal_2024}, or MnTe~\cite{Mazin_2023}, enabling the intrinsic symmetry breaking required for topological superconductivity and the SDE without external magnetic fields.
In this broader context, the proposed altermagnetic Shiba chain provides the first material-specific solid-state route to a tunable topological FF phase, previously explored mainly in spin–orbit–coupled cold-atom systems~\cite{Zhang2013topological,Qu2013topological}, but now achievable in a magnetic-field-free architecture. Building on this foundation, several promising future directions emerge. Extending our framework to two-dimensional Shiba lattices coupled to $d$-wave altermagnets may uncover richer topological phases and enhanced SDE, driven by the more intricate momentum-space structure of altermagnetism. In particular, it would be compelling to investigate whether such systems can intrinsically realize a perfect superconducting diode effect (recently demonstrated in altermagnetic superconductors under applied magnetic fields~\cite{Debmalya_2025}) without the need for any external field.
Additionally, exploring alternative altermagnetic pairing symmetries, such as $p$- or $g$-wave order~\cite{Tanaka_2025,Ezawa_2025}, may uncover novel mechanisms for nonreciprocal superconductivity and expand the accessible material landscape. Altogether, our results establish altermagnet–superconductor hybrids as a versatile platform for integrating topological quantum computation and low-dissipation superconducting electronics within a single, field-free device architecture—paving the way toward scalable quantum technologies.

%~~~~~~~~~~~~~~~~~~~~~~~~~~~~~~~~~~~~~~~~~
\section{Acknowledgments} 
%~~~~~~~~~~~~~~~~~~~~~~~~~~~~~~~~~~~~~~~~~
SKG acknowledges financial support from Anusandhan National Research Foundation (ANRF) erstwhile Science and Engineering Research Board (SERB), Government of India via the Startup Research Grant: SRG/2023/000934 and from IIT Kanpur via the Initiation Grant (IITK/PHY/2022116). DS and SKG utilized the \textit{Andromeda} server at IIT Kanpur for numerical calculations.

\bibliography{main_refs}

\clearpage
\onecolumngrid 

\section*{Supplementary material}
% \section*{Supplementary material to ``Field-free Superconducting Diode Effect and Topological Fulde-Ferrell Superconductivity in Altermagnetic Shiba Chains''}

\setcounter{section}{0} % restart numbering
\renewcommand{\thesection}{S\arabic{section}} % S1, S2, ...

\setcounter{figure}{0} % restart numbering
\renewcommand{\thefigure}{S\arabic{figure}} % S1, S2, ...

\setcounter{equation}{0} % restart numbering
\renewcommand{\theequation}{S\arabic{equation}} % S1, S2, ...

% \title{Supplementary material to \\ ``Field-free Superconducting Diode Effect and Topological Fulde-Ferrell Superconductivity in Altermagnetic Shiba Chains''}

This supplementary text provides additional technical details supporting the results in the main text. In \sect{sec:momentum_space_ham}, we present the derivation of the momentum-space Hamiltonian for the Shiba chain. In \sect{sec:momentum_space_sde}, we analyze the superconducting diode effect within the momentum-space framework. Finally, in \sect{sec:topology_canonical}, we explore the topological properties of a conical Shiba chain proximitized by a $d$-wave altermagnet under open boundary conditions.

\section{Derivation of the momentum space Hamiltonian}\label{sec:momentum_space_ham}

The BdG Hamiltonian in real space in the low-energy continuum limit for our proposed setup, expressed in the Nambu basis $\Psi(x) = [c_\ua (x), c_\da (x), c_{\ua}^\dagger (x),c_{\da}^\dagger (x)]^T$, is given by:

\bea
    \mathcal{H} &=& \frac{1}{2} \int dx \Psi^\dagger(x) \mathcal{H}_{BdG} \Psi(x) + \epsilon_0\,, \non \\
    \mathcal{H}_{BdG} &=&
    \begin{bmatrix}
        h (x) & \Delta (x) \\
        -\Delta^* (x) & -h^*(x)
    \end{bmatrix} \,, \non \\
    h(x) &=&  \bigg(-\frac{\hbar^2}{2m}\partial_{x}^2 - \mu \bigg) + J \mathbf{S} (x) \cdot \boldsymbol{\sigma} + J_A \bigg(1 + \frac{1}{2} \partial_{x}^2 \bigg) \sigma_z \,,
\eea

\noindent where, $\pmb{\sigma}$ denotes the vector of Pauli matrices acting in spin space. The pairing term $\Delta(x) = -i \sigma_y \Delta_0 e^{i q x}$ corresponds to an $s$-wave Fulde-Ferrell (FF) order parameter, with $q$ representing the finite momentum of the Cooper pairs. A Shiba chain made of atoms with a spin texture can be mapped onto a Rashba nanowire under a uniform magnetic field via a local, spin-dependent gauge transformation~\cite{Hess_2022}. Specifically, under the unitary transformation~\cite{Hess_2022,Pritam2024}
\bea
    \tilde{h} = U^\dagger h U \;\;\text{with,}\hspace{4mm} U=e^{-\frac{i}{2} \phi(x) \sigma_z}\,,
\eea
we obtain the low energy Hamiltonian as:
\beq
    \tilde{h} = \left(-\frac{\hbar^2}{2m} \mathcal{D}_{x}^2 -\mu\right) + J \sin \theta \sigma_x + J \cos \theta \sigma_z + J_A \left(1 + \frac{1}{2} \mathcal{D}_{x}^2 \right) \sigma_z\,,
\eeq
\noindent where the covariant derivative is defined as $\mathcal{D}_x = \partial_x - \frac{i}{2} \partial_x \phi(x), \sigma_z$. Assuming a constant pitch vector, such that $g = \partial_x \phi(x)$ is uniform, a Fourier transformation yields the corresponding continuum Hamiltonian in momentum space to be:
\beq
    \tilde{h}_k = \xi_{k,\tilde{g}} - \mu - \frac{\hbar^2}{m} \tilde{g} k \sigma_z + J \sin \theta \sigma_x + J \cos \theta \sigma_z + J_A \bigg(1 - \frac{1}{2} (k^2 + \tilde{g}^2) \bigg) \sigma_z + J_A \tilde{g} k\,,
\eeq
\noindent where $\xi_{k,\tilde{g}} = \frac{\hbar^2}{2m} (k^2 + \tilde{g}^2)$, with $\tilde{g} = g/2$. The second term acts as an effective spin-orbit coupling (SOC), emerging from the spin texture inherent in our model. In contrast to conventional SOC, which originates from intrinsic relativistic effects, this term arises due to the spatial variation in the spin configuration.

%%%%%%%%%%%%%%%%%%%%%%%%%%%%%%%%%%%%%%%%%%%%%%%%%%%%%%%%%%%%%%%%%%%%%%%%%%%%%%%%%%%%%%%%%%%%%%%%%%%%%%%%%%%%%%%%%%%%%%%%%%%%%%%%%%%%%%%%%%%%
\begin{figure*}[!htb]
\centering 
\includegraphics[width=\linewidth]{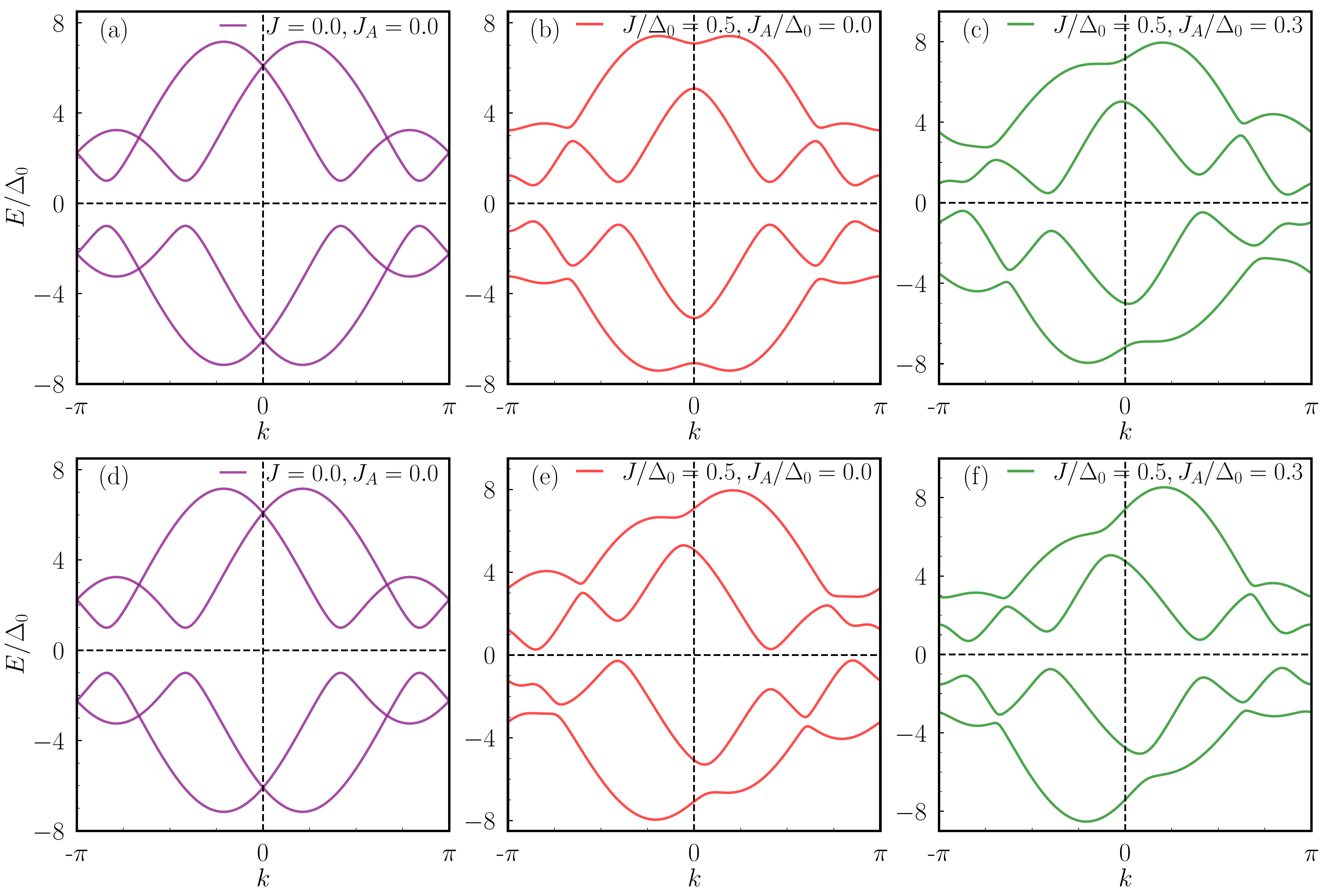}
\caption{\textbf{Bogoliubov quasiparticle dispersions:}  Quasiparticle energy spectra for a one-dimensional Shiba chain with (a–c) helical and (d–f) conical spin textures in proximity to a $d$-wave altermagnet. Panels (a) and (d) show the spectrum in the absence of both magnetic exchange field and altermagnetic strength ($J = J_A = 0$). Panels (b) and (e) display the effect of exchange field alone ($J/\Delta_0 = 0.5$, $J_A = 0$), while panels (c) and (f) include both exchange interaction and altermagnetic strength ($J/\Delta_0 = 0.5$, $J_A/\Delta_0 = 0.3$). The supporting model parameters are: ($t/\Delta_0$, $g$, $\mu/\Delta_0$, $q$) = ($1.0$, $\pi/ 2$, $t(2 + g^2/4) + 1.0$, $0$).}
\label{fig:dispersion}
\end{figure*}
%%%%%%%%%%%%%%%%%%%%%%%%%%%%%%%%%%%%%%%%%%%%%%%%%%%%%%%%%%%%%%%%%%%%%%%%%%%%%%%%%%%%%%%%%%%%%%%%%%%%%%%%%%%%%%%%%%%%%%%%%%%%%%%%%%%%%%%%%%%%

%%%%%%%%%%%%%%%%%%%%%%%%%%%%%%%%%%%%%%%%%%%%%%%%%%%%%%%%%%%%%%%%%%%%%%%%%%%%%%%%%%%%%%%%%%%%%%%%%%%%%%%%%%%%%%%%%%%%%%%%%%%%%%%%%%%%%%%%%%%%%
\begin{figure}[t]
\centering 
\includegraphics[width=0.8\textwidth]{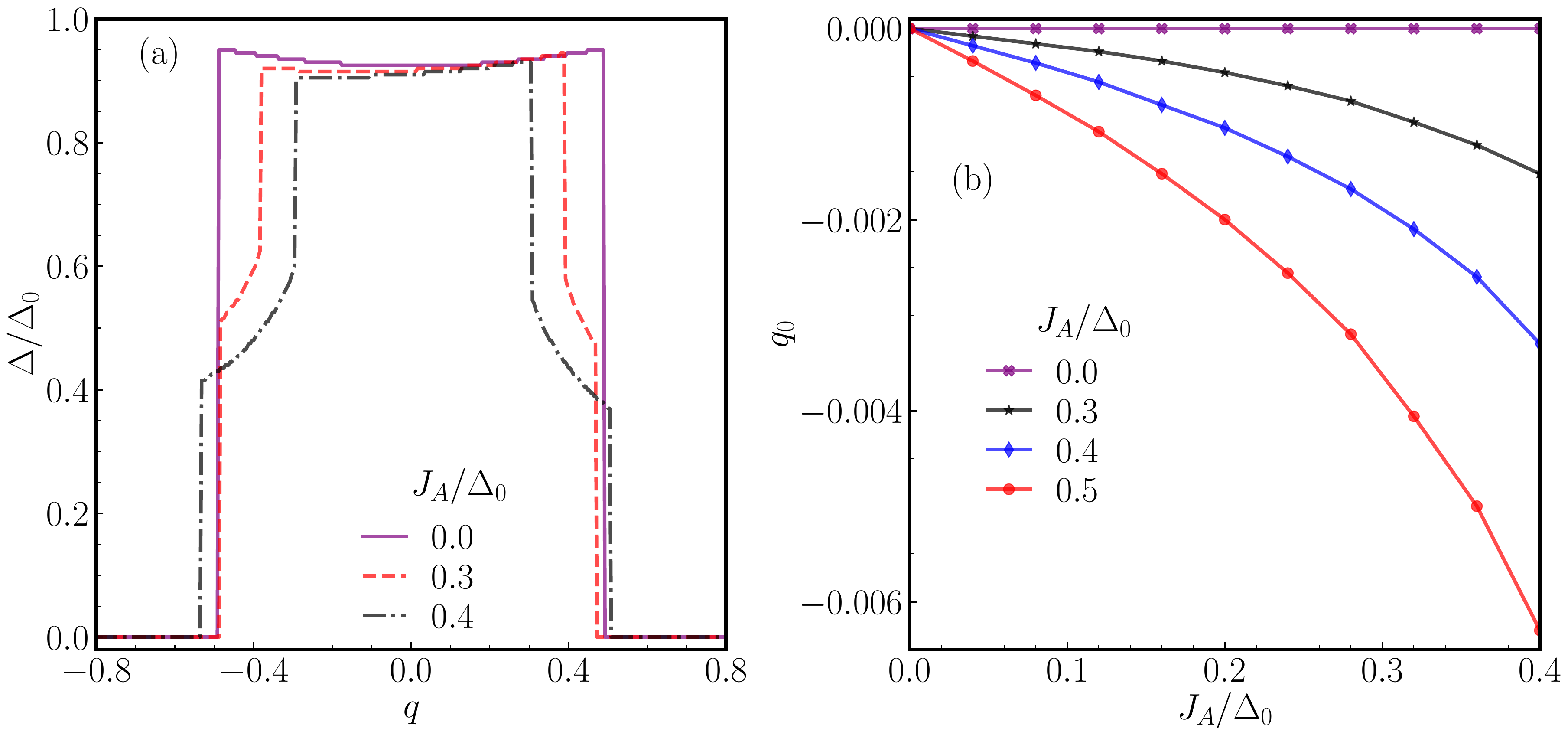}
\caption{\textbf{FF ground state in momentum space:} (a) The self-consistent superconducting gap $\Delta(q)$ is shown as a function of the Cooper pair momentum $q$ for an exchange field $J/\Delta_0 = 0.5$, where $\Delta_0$ represents the superconducting gap of a 3D $s$-wave superconductor. (b) The optimal momentum $q_0$ corresponding to the FF ground state is plotted as a function of the proximity-induced altermagnetic exchange strength $J_A$. The model parameters used are: ($t/\Delta_0$, $g$, $U/\Delta_0$, $\mu/\Delta_0$, $(\beta~\Delta_0)^{-1}$) = ($1.0$, $2 \pi/3$, $0.2985$, $t(2 + g^2/4) + 1.0$, $0.01$).}   
\label{fig:fflo_k_space}
\end{figure}
%%%%%%%%%%%%%%%%%%%%%%%%%%%%%%%%%%%%%%%%%%%%%%%%%%%%%%%%%%%%%%%%%%%%%%%%%%%%%%%%%%%%%%%%%%%%%%%%%%%%%%%%%%%%%%%%%%%%%%%%%%%%%%%%%%%%%%%%%%%%

To derive the lattice counterpart of the continuum Hamiltonian above, we employ the standard lattice regularization replacements $k \rightarrow \sin k$ and $1 - \frac{k^2}{2} \rightarrow \cos k$, resulting in:
\beq
    h_k = \eps_{k,\tilde{g}} - t~g \sin k \, \sigma_z + J \sin \theta \sigma_x + J \cos \theta \sigma_z + J_A \bigg(\cos k - \frac{\tilde{g}^2}{2}\bigg) \sigma_z + J_A \tilde{g} \sin k\,, \label{eq:k_space_hamiltonian}
\eeq
\noindent where $\eps_{k,\tilde{g}} = t(2 - 2\cos k) + t\tilde{g}^2 - \mu$ with $t=\frac{\hbar^2}{2m}$. For simplicity, we set $\hbar = m = 1$ throughout our analysis. In general, the presence of supercurrents can lead to a superconducting state characterized by finite-momentum Cooper pairing, known as the FF state. The mean field BDG hamiltonian in the basis $\psi(k,q)=\left( c_{k+ q/2,\ua}, c_{k+ q/2,\da}, c_{-k+ q/2,\da}^\dagger, -c_{-k+ q/2,\ua}^\dagger \right)^T$ reads as:

\bea
\mathcal{H}_{MF} &=& \frac{1}{2} \sum_{k} \psi^\dagger(k,q) H(k, q) \psi(k,q) + \epsilon_0 \,, \non\\
&=& \sum_{n,k} \left(E_{n,k} \gamma^\dagger_{n,k} \gamma_{n,k} - \frac{E_{n,k}}{2} \right) + \epsilon_0 \,, \non 
\eea
\vspace{-0.3cm}
\beq
\text{\rm with,}\;\;H(k,q) =
    \begin{bmatrix}
	h_{k+q/2} & -i \sigma_y \Delta \\
	i \sigma_y \Delta & -h^*_{-k + q/2}
    \end{bmatrix} \label{eq:k_space_model}\,. 
\eeq

\noindent As in the real-space lattice case, we determine the Cooper pair momentum $q_0$ by optimizing the condensation energy using the self-consistent solution for $\Delta(q)$ [Eq.~(5) of main text]. The BCS gap $\Delta_0$ refers to that of a 3D $s$-wave superconductor, and the Hubbard interaction strength $U$ is chosen to ensure the system remains in the weak-coupling BCS regime, with $\Delta_0 \sim 1$ meV.

We analyze the Bogoliubov quasiparticle spectra of the Hamiltonian in \eqn{eq:k_space_model}, as shown in \fig{fig:dispersion}, to investigate the origin of spectral asymmetry in helical and conical Shiba chains proximitized by a $d$-wave altermagnet. In the absence of both the exchange interaction ($J$) and the altermagnetic proximity ($J_A$), the spectra remain symmetric for both helical [\fig{fig:dispersion}(a)] and conical [\fig{fig:dispersion}(d)] spin textures, reflecting preserved inversion and time-reversal symmetries. For the helical texture, turning on $J$ alone [\fig{fig:dispersion}(b)] does not induce asymmetry in the spectrum, as the helical configuration does not break inversion symmetry on its own. Spectral asymmetry appears only when both $J$ and $J_A$ are nonzero [\fig{fig:dispersion}(c)], demonstrating that the altermagnet plays an essential role in breaking inversion symmetry for the helical chain. In contrast, for the conical spin texture, the inclusion of a finite $z$-component inherently breaks inversion symmetry. As a result, $J$ alone is sufficient to generate asymmetric quasiparticle bands [\fig{fig:dispersion}(e)], and introducing $J_A$ further reduces this asymmetry [\fig{fig:dispersion}(f)]. This comparison highlights the distinct symmetry-breaking mechanisms that govern spectral asymmetry and nonreciprocal transport in the two spin textures.

\fig{fig:fflo_k_space} shows the variation of the self-consistent superconducting order parameter $\Delta(q)$ and the optimal Cooper pair momentum $q_0$ in the FF ground state of a helical Shiba chain with in-plane helical spin texture, as a function of the induced altermagnetic strength $J_A$. As seen in \fig{fig:fflo_k_space}(a), the superconducting gap vanishes beyond critical values of the Cooper pair momentum, either above a positive threshold $q = q_c^+$ or below a negative threshold $q = q_c^-$. \fig{fig:fflo_k_space}(b) reveals that $q_0$ increases approximately linearly with $J_A$ for $J_A \ll \Delta_0$, at a fixed exchange coupling $J$.

%%%%%%%%%%%%%%%%%%%%%%%%%%%%%%%%%%%%%%%%%%%%%%%%%%%%%%%%%%%%%%%%%%%%%%%%%%%%%%%%%%%%%%%%%%%%%%%%%%%%%%%%%%%%%%%%%%%%%%%%%%%%%%%%%%%%%%%%%%%%
\begin{figure*}[!htb]
\centering 
\includegraphics[width=\linewidth]{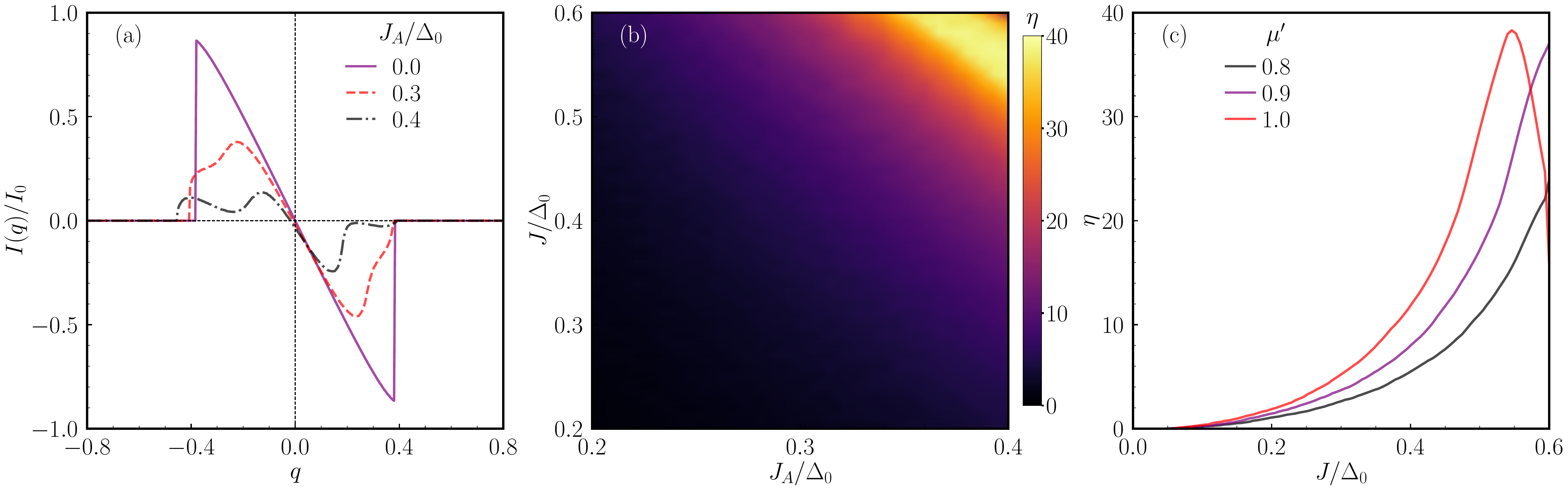}
\caption{\textbf{Superconducting diode effect for helical spin texture in momentum space:} (a) The supercurrent density ($I$) is plotted as a function of $q$ for $J/\Delta_0 = 0.5$. Nonreciprocal behavior, indicated by $I(q) \ne -I(-q)$, appears only when both $J$ and $J_A$ are finite. This leads to asymmetric critical currents, $I_c^+ \ne |I_c^-|$, and a nonzero diode efficiency, $\eta \ne 0$. The symmetric critical current in the absence of altermagnetic exchange is given by $I_0 \equiv |I_c^+(J_A = 0)| = |I_c^-(J_A = 0)|$. (b) The diode efficiency $\eta$ is mapped in the $J$–$J_A$ plane for a chemical potential $\mu = t(2 + g^2/4) + 1.0$. (c) The dependence of $\eta$ on exchange coupling is shown for different values of the effective chemical potential, defined as $\mu^\prime = \mu - t(2 + g^2/4)$. The other model parameters are: ($t/\Delta_0$, $g$, $U/\Delta_0$, $(\beta~\Delta_0)^{-1}$) = ($1.0$, $2 \pi/3$, $0.279$, $0.1$).}
\label{fig:kspace_sde}
\end{figure*}
%%%%%%%%%%%%%%%%%%%%%%%%%%%%%%%%%%%%%%%%%%%%%%%%%%%%%%%%%%%%%%%%%%%%%%%%%%%%%%%%%%%%%%%%%%%%%%%%%%%%%%%%%%%%%%%%%%%%%%%%%%%%%%%%%%%%%%%%%%%%

%-------------------------------------------------------------
\section{Superconducting diode effect in momentum space}\label{sec:momentum_space_sde}
%-------------------------------------------------------------

The momentum-space Hamiltonian in Eq.\eqref{eq:k_space_hamiltonian} breaks both inversion and time-reversal symmetry, enabling the superconducting diode effect (SDE). To examine this effect in an in-plane helical Shiba chain, we analyze the supercurrent density $I(q)$, which shows a clear asymmetry when both the exchange coupling $J$ and the induced altermagnetic strength $J_A$ are finite, as shown in \fig{fig:kspace_sde}(a). \fig{fig:kspace_sde}(b) presents the diode efficiency $\eta$ as a function of $J$ and $J_A$, revealing that the efficiency increases with both parameters and reaches a maximum of around $\sim 40 \%$. In \fig{fig:kspace_sde}(c), we plot $\eta$ as a function of $J$ for different values of the effective chemical potential $(\mu - t (2 + g^2/4))$. The efficiency grows monotonically with $J$, reflecting the enhanced asymmetry in the quasiparticle spectrum. The maximum efficiency increases as one increases the magnitude of chemical potential.

%%%%%%%%%%%%%%%%%%%%%%%%%%%%%%%%%%%%%%%%%%%%%%%%%%%%%%%%%%%%%%%%%%%%%%%%%%%%%%%%%%%%%%%%%%%%%%%%%%%%%%%%%%%%%%%%%%%%%%%%%%%%%%%%%%%%%%%%%%

\begin{figure}[b]
\centering 
\includegraphics[width=0.6\textwidth]{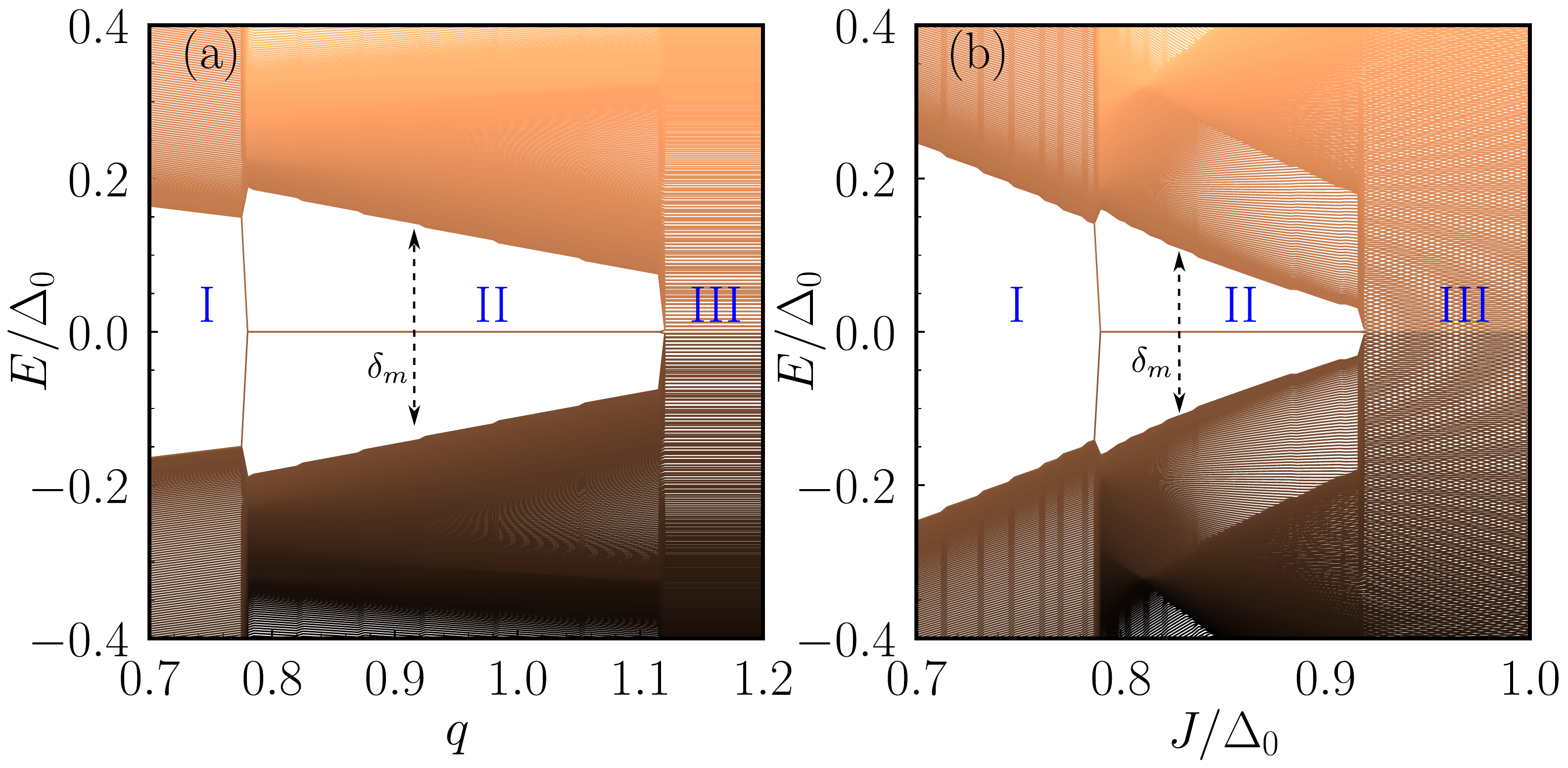}
\caption{\textbf{Bogoliubov quasiparticle spectrum with OBC for a conical Shiba chain} (a) The eigenvalue spectrum is shown as a function of Cooper pair momentum $q$, incorporating the self-consistently obtained pairing gap $\Delta(q)$ for $J/\Delta_0 = 0.8$ and altermagnetic strength $J_A/\Delta_0 = 0.3$. (b) The spectrum is plotted as a function of the exchange field $J$ for fixed $q = 0.9$ and $J_A/\Delta_0 = 0.3$. The spectrum reveals three distinct regions: (1) a trivially gapped phase, (2) a topologically nontrivial gapped phase hosting MZMs, protected by an effective minigap $\delta_m$, and (3) a normal phase where the superconducting gap $\Delta(q)$ vanishes. The chosen model parameters are ($t/\Delta_0$, $g$, $U/\Delta_0$, $\mu/\Delta_0$, $(\beta~\Delta_0)^{-1}$) = ($0.5$, $\pi/2$, $1.38$, $1.0$, $0.01$). All numerical results are obtained for a finite lattice of $400$ sites employing OBC in the BdG lattice Hamiltonian.}   
\label{fig:spectrum_obc_canonical}
\end{figure}

%%%%%%%%%%%%%%%%%%%%%%%%%%%%%%%%%%%%%%%%%%%%%%%%%%%%%%%%%%%%%%%%%%%%%%%%%%%%%%%%%%%%%%%%%%%%%%%%%%%%%%%%%%%%%%%%%%%%%%%%%%%%%%%%%%%%%%%%%%

%%%%%%%%%%%%%%%%%%%%%%%%%%%%%%%%%%%%%%%%%%%%%%%%%%%%%%%%%%%%%%%%%%%%%%%%%%%%%%%%%%%%%%%%%%%%%%%%%%%%%%%%%%%%%%%%%%%%%%%%%%%%%%%%%%%%%%%%%%
\begin{figure*}[!htb]
\centering 
\includegraphics[width=\linewidth]{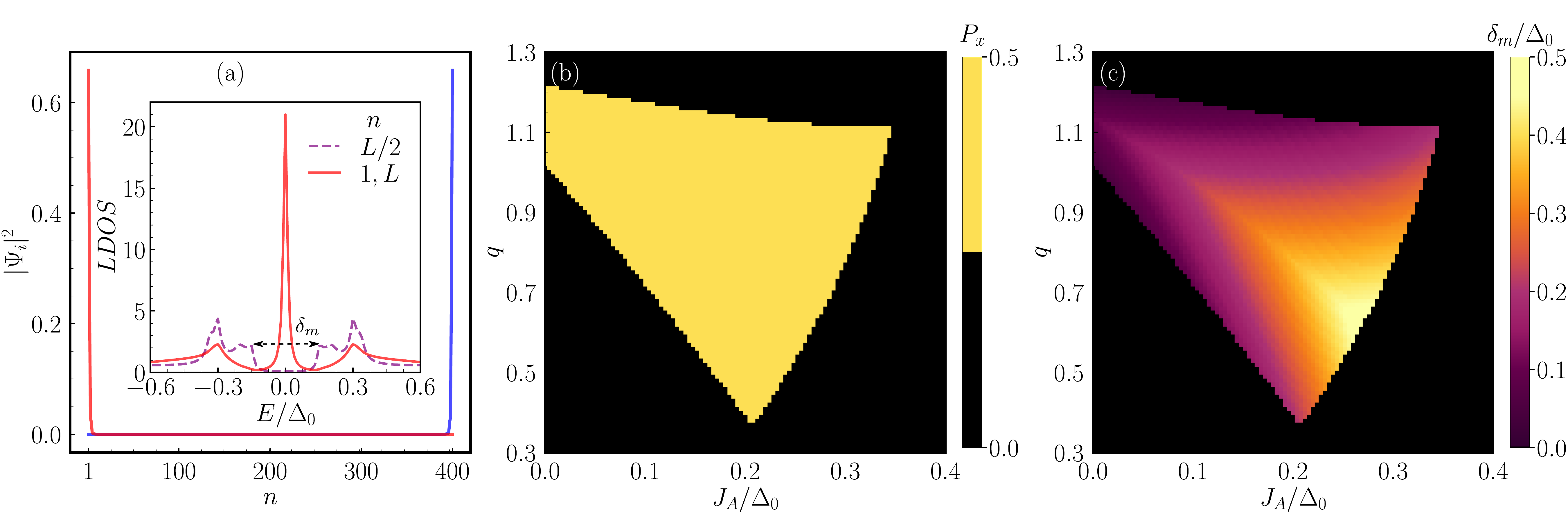}
\caption{\textbf{Topological signatures of MZMs for a conical Shiba chain:} (a) The site-resolved normalized probability density $|\Psi_i|^2$ of the two Majorana zero modes (MZMs) is shown for fixed $q = -0.8$ and $J_A/\Delta_0 = 0.3$, revealing their strong localization at the ends of the chain. The inset displays the energy-resolved local density of states (LDOS) at the end (solid red line) and the middle (dashed blue line) of the chain. (b) and (c): The profiles of the effective minigap $\delta_m$ and the topological invariant $P_x$ (bulk dipole moment) are presented in the $J_A$–$q$ parameter space. The model parameters used are ($t/\Delta_0$, $g$, $U/\Delta_0$, $\mu/\Delta_0$, $(\beta~\Delta_0)^{-1}$, $J/\Delta_0$) = ($0.5$, $\pi/2$, $1.38$, $1.0$, $0.01$, $0.65$).}   
\label{fig:topology_conical}
\end{figure*}
%%%%%%%%%%%%%%%%%%%%%%%%%%%%%%%%%%%%%%%%%%%%%%%%%%%%%%%%%%%%%%%%%%%%%%%%%%%%%%%%%%%%%%%%%%%%%%%%%%%%%%%%%%%%%%%%%%%%%%%%%%%%%%%%%%%%%%%%%%

%-------------------------------------------------------------
\section{Topological Features of the Conical Shiba Chain proximitized with a $d$-Wave Altermagnet}\label{sec:topology_canonical}
%-------------------------------------------------------------

To reveal the topological features of a conical Shiba chain proximitized by a $d$-wave altermagnet, we examine the BdG quasiparticle energy spectrum under open boundary conditions as a function of the Cooper pair momentum $q$ and the exchange coupling $J$, as shown in \fig{fig:spectrum_obc_canonical}. The spectra exhibit three distinct regimes: Regime (I) corresponds to a trivially gapped superconducting phase; Regime (II) features doubly degenerate zero-energy in-gap states, indicating a topological superconducting phase with Majorana zero modes (MZMs) driven by finite-momentum pairing; and Regime (III) represents a gapless normal phase where superconductivity is destroyed and the order parameter $\Delta$ vanishes.

To identify the emergence of MZMs in the conical Shiba chain proximitized by a $d$-wave altermagnet, we first analyze the LDOS in \fig{fig:topology_conical}(a), where the chain ends exhibit a pronounced zero-energy peak separated from the bulk by a finite minigap $\delta_m$, confirming strong end localization of MZMs, while the chain center shows only subgap Yu–Shiba–Rusinov (YSR) bands. To quantify the bulk topology, we compute the polarization $P_x$ using the self-consistent pairing amplitude $\Delta$, where $P_x = 0.5$ signals a topological phase and $P_x = 0.0$ corresponds to a trivial one. The resulting phase diagrams in \fig{fig:topology_conical}(b,c) show that regions with $P_x = 0.5$ coincide with a finite minigap, establishing a robust topological superconducting phase in the conical spin texture.

\end{document}